\documentclass[a4paper,fleqn,usenatbib,useAMS]{mnras}
\usepackage{graphicx}
\usepackage{epstopdf}
\usepackage{pdflscape}
\usepackage{subfig}
\usepackage{float}
\usepackage{xcolor}
\usepackage{rotating}
\usepackage{pdflscape}

\title[Inverted Spectrum Extragalactic Radio Sources]{GMRT observations of a first sample of \lq Extremely Inverted Spectrum Extragalactic Radio Sources (EISERS)\rq~candidates in the Northern sky}

\author[Mukul-Mhaskey et al.]
  {Mukul Mhaskey$^{1,2}$\thanks{E-mail:mhaskeymukul@gmail.com},
   Gopal-Krishna$^{3}$,
   Surajit Paul$^{1}$,
   Pratik Dabhade$^{4,5}$,    
\newauthor
   Sameer Salunkhe$^{1}$,
   Shubham Bhagat$^{1}$ and
   Abhijit Bendre$^{4}$
\\
$^{1}$Department of Physics, Savitribai Phule Pune University, Ganeshkhind, Pune 411007, India\\
$^{2}$Th\"uringer Landessternwarte, Sternwarte 5, D-07778 Tautenburg, Germany\\
$^{3}$Aryabhatta Research Institute of Observational Sciences (ARIES), Manora Peak, Nainital$-$263129, India\\
$^{4}$Inter University Centre for Astronomy and Astrophysics (IUCAA), Pune 411007, India\\
$^{5}$Leiden Observatory, Leiden University, Niels Bohrweg 2, 2333 CA, Leiden, Netherlands}
\pubyear{2018}

\begin{document}
\label{firstpage}
\pagerange{\pageref{firstpage}--\pageref{lastpage}}
\maketitle

\begin{abstract}
We present an extension of our search for `Extremely Inverted Spectrum Extragalactic Radio Sources' (EISERS) to the northern 
celestial hemisphere. With an inverted radio spectrum of slope $\alpha$ $>$ $+$2.5, these rare sources would either require a non-standard 
particle acceleration mechanism (in the framework of synchrotron self-absorption hypothesis), or a severe free-free absorption 
which attenuates practically all of their synchrotron radiation at metre wavelengths. By applying a sequence of selection filters,
a list of 15 EISERS candidates is extracted out by comparing two large-sky radio surveys, WENSS (325 MHz) and TGSS-ADR1 (150 MHz), which 
overlap across 1.03$\pi$ steradian of the sky. Here we report quasi-simultaneous GMRT observations of these 15 EISERS candidates at 150 MHz and 
325 MHz, in an attempt to accurately define their spectra below the turnover frequency. Out of the 15 candidates observed, two are confirmed as EISERS, since the slope of the inverted spectrum between these two frequencies is found to be significantly larger than the critical value $\alpha_c$ = +2.5: the theoretical limit for the standard case of synchrotron self-absorption (SSA). For another 3 sources, the spectral slope is close to, or just above the critical value $\alpha_c$. Nine of the sources have GPS type radio spectra. The parsec-scale radio structural information available for the sample is also summarised. 
 
\end{abstract}

\begin{keywords}
radiation mechanisms: non thermal -- galaxies: ISM --
galaxies: jets -- galaxies: nuclei -- quasars: general -- 
radio continuum: galaxies
\end{keywords}

\section{Introduction}\label{section1}
Sensitive large-area radio surveys, such as the FIRST \citep{Helfand2015} and NVSS \citep{Condon1998}, combined with their 
optical counterpart, SDSS \citep{Alam2015} have enabled the assembly of large samples for detailed studies of different 
categories of extragalactic radio sources \citep{Best2005, Coziol2017}. In recent years, such samples have been increasingly 
used for finding not only the rare types of radio sources, such as X-shaped radio sources \citep[][and references therein]{Cheung2013, Roberts2018, GK2012} 
, giant radio galaxies \citep{Dabhade2017}, recurrently active radio galaxies \citep{Kuzmicz2017} and HYMORS \citep{Kapinska2017, Gopal-Krishna2000}, but also species like `Compact Symmetric Objects' \citep {Gugliucci2005, Wilkinson1994, Conway1994, Phillips1982} and FR0 radio galaxies \citep{Baldi2015} whose abundance had been scarcely recognized. 
A major advance in recent times is the availability of large-area surveys at metre wavelengths, such as the TGSS-ADR1 
\citep{Intema2017}, LoTSS \citep{Shimwell2017, Shimwell2019} and the GLEAM survey \citep{Wayth2015, Hurley-Walker2017}. These go significantly deeper than their precursors, 
the Westerbork surveys, WENSS at 325 MHz \citep{deBruyn2000} and WISH at 352 MHz \citep{DeBreuck2002}, as well as the VLA 
74 MHz survey \citep{Lane2014}. 

A few years ago, we combined the TIFR GMRT SKY SURVEY (TGSS/DR5) at 150 MHz with the 352 MHz WISH survey, to search for extragalactic radio 
sources whose radio spectrum shows an ultra-sharp turnover, such that the slope of the inverted spectrum is greater than the 
critical value $\alpha_c$ $=$ $+$2.5: the theoretical limit for the standard case of synchrotron self-absorption (SSA)
\citep{GopalKrishna2014} (henceforth Paper I). This limit cannot be violated even by a perfectly homogeneous source of incoherent 
synchrotron radiation arising from relativistic electrons whose energy distribution has the canonical (i.e., power-law) shape 
\citep{Slish1963,Scheuer1968,Rybicki1979}. Consequently, any such sources would either be having a non-standard electron energy 
distribution \citep{Rees1967}, or suffer a severe free-free absorption (FFA) at metre wavelengths, as discussed in Paper I where 
they were christened `Extremely Inverted Spectrum Extragalactic Radio Sources' (EISERS). The 7 EISERS candidates reported in 
Paper I were subsequently followed up with the Giant Metrewave Radio Telescope - GMRT \citep{Swarup1991}, by making quasi-simultaneous 
observations at 150, 325, 610 and 1400 MHz \citep[][henceforth Paper II]{Mhaskey2019}. This and a similarly targeted independent 
program, based on the GLEAM survey in which flux densities are measured simultaneously at several narrow bands in the 72-231 MHz 
range, although with a modest sensitivity and resolution \citep{Callingham2017}, have together been able to find just a couple 
of EISERS, which testifies to extreme rarity of such radio sources. The aim of the present study is to extend our search for EISERS 
to the northern sky. This was done by combining the available two large-area, high-resolution radio surveys, namely the recently published TGSS-ADR1 at 150 MHz \citep{Intema2017} and the pre-existing WENSS at 325 MHz \citep{deBruyn2000}. The choice of these two 
large-area radio surveys was motivated by their fairly high sensitivity (typical rms $<$ 10 mJy/beam) and sub-arcminute resolution
at metre wavelengths. The EISERS candidates in the Northern sample were then observed quasi-simultaneously with the GMRT.

In Section \ref{section2} we describe the selection procedure for the 15 EISERS candidates. Section \ref{section3} contains the details of their quasi-simultaneous radio observations and the analysis procedure. Notes on individual sources are given in Section \ref{section4}. The results of spectral modelling on the EISERS candidates are discussed in Section \ref{section5}. 
The following Sections \ref{section6} and \ref{section7} contain a brief discussion on EISERS and the conclusions, respectively.

\section{Selection of EISERS candidates in the northern sky}\label{section2}
The two main radio surveys (Section \ref{section1}), viz., the TGSS-ADR1 (150 MHz) and WENSS (325 MHz), overlap across a region 
of about 1.03$\pi$ steradian ($\sim$1/4th of the entire sky). This overlapping region contains 229420 radio sources in the WENSS catalogue covering the declination range north of +28$^\circ$. Out of this extensive source list, we first extracted a subset containing 35064 sources which belong to the morphological type \lq S\rq~(i.e., single, as per the WENSS catalogue) and are stronger than 150 mJy at 325 MHz. For each of these shortlisted WENSS sources, we then looked for a TGSS-ADR1 counterpart, within a search radius 
of 20 arcsec. Thus, counterparts were found for 33707 (out of the 35064 sources) of the WENSS sources. For 1357 WENSS source no counterparts were found in TGSS-ADR1. The median separation between the positions of sources from the WENSS and TGSS-ADR1 is found to be 2.10$\pm$0.02 arcsec, which is consistent with the quoted positional uncertainties for the relatively weak sources in the two surveys (rms $\sim$ 1 arcsec for WENSS and $\sim$ 2 arcsec for TGSS-ADR1). The spectral index, $\alpha$, between 150 and 325 MHz, was calculated for all the sources. For the WENSS sources without a counterpart (1357) in the TGSS-ADR1, a 5-$\sigma$ upper limit to the flux at 150 MHz was used to estimate the spectral index. Here $\sigma$ is the image rms for individual sources in the TGSS-ADR1.    

The final list of EISERS candidates satisfy the following criteria: (i) structural type \lq S\rq~and flux density $>$ 150 mJy at 325 MHz, (ii) omission of sources found to lie within 10$^{\circ}$ of the galactic plane, or listed as H-II regions in the NASA Extragalactic Database (NED)\footnote{https://ned.ipac.caltech.edu/} (iii) an unambiguous detection in the WENSS (325 MHz) and NVSS (1.4 GHz) (based on visual inspection of the respective radio images) and (iv) $\alpha$ (150-325 MHz) $>$ $+$2.75, in case of sources detected in TGSS-ADR1 or $\alpha$ (150-325 MHz) $>$ $+$2.5, in case of sources not detected in TGSS-ADR1. A conservative threshold value of +2.75 was adopted for sources detected in TGSS-ADR1 in order to make an allowance for flux variability, which could be significant in case of compact sources (see Section \ref{section6}). The final list contains 15 EISERS candidates, which is presented in Tables~\ref{table:obs-log} and \ref{table:spec-prop} along with their flux densities taken from the literature, as well as those determined from the new quasi-simultaneous observations reported here (Section \ref{section3}). The radio contour images taken with the uGMRT at 150 and 325 MHz of the two best candidates are displayed in Figures~\ref{fig:J1326} and \ref{fig:J1658}. Radio spectra of the 15 sources, based on the data provided in Table~\ref{table:spec-prop} are shown in Figure~\ref{fig:spec_all}. Table~\ref{table:summary} is the summary table that contains the computed spectral index $\alpha_{thick}$ (150-325 MHz) for the optically thick part of the spectrum. $\alpha_{thick}$$>$+2.5 is the principal marker for EISERS. Table~\ref{table:summary} also contains the spectral index, $\alpha_{thin}$ for the optically thin part of the spectrum, along with the 5 GHz radio power for each source. For sources with unknown redshift, a redshift z = 1 has been assumed. 
   
\section{Radio Observations and Analysis}\label{section3}
The need for quasi-simultaneous radio observations at frequencies below the spectral turnover is underscored by the fact that the two radio surveys (TGSS-ADR1 and WENSS) used for computing the spectral slopes of these compact radio sources had been made nearly a decade apart. The long time interval could then have introduced significant uncertainty due to flux variability expected from 
refractive interstellar scintillation at such low frequencies, e.g. \citep{Bell2019} (Section \ref{section6}). It may also be noted that the WENSS \citep{Rengelink1997} is known to be off the flux-density scales defined by Roger, Costain \& Bridle \citep[RCB,][]{Roger1973} and by \citet{Baars1977} by over 10\% \citep[see,][]{Hardcastle2016}. Therefore, the previous spectral index estimates could be substantially in error due to the combined effect of measurement uncertainty and the calibration uncertainties of the WENSS and TGSS-ADR1 flux densities. This may account for the substantial differences found in some of the cases, between the WENSS flux density at 325 MHz and the present GMRT measurements at the same frequency (Section \ref{section4}). For our sample of 15 EISERS candidates, the present GMRT observations have yielded the data with the highest sensitivity and resolution currently available at 150 and 325 MHz (Table~\ref{table:obs-log}). Moreover, the availability of two data points at well-spaced frequencies in their highly opaque spectral region raises the confidence in quantifying the steepness of the spectral turnover. 

\begin{table*}
\caption{The uGMRT observation log and the basic parameters of the radio maps.\\}
\label{table:obs-log}
\begin{tabular}{cccccccc}\\
\hline

\multicolumn{1}{c}{Source} & \multicolumn{1}{c}{Coordinates} & Obs. date & \multicolumn{1}{c}{No. of} & \multicolumn{1}{c}{Total} 
& \multicolumn{2}{c}{Synthesised beam}  &  \multicolumn{1}{c}{Map rms} \\
\multicolumn{1}{c}{} &\multicolumn{1}{c}{(J2000)} &  & \multicolumn{1}{c}{snapshots} & \multicolumn{1}{c}{integration} 
& \multicolumn{1}{c}{FWHM} & \multicolumn{1}{c}{PA} & \multicolumn{1}{c}{noise} \\
\multicolumn{1}{c}{} & &  & \multicolumn{1}{c}{} & \multicolumn{1}{c}{(min)} 
& \multicolumn{1}{c}{(arc sec)} & \multicolumn{1}{c}{(deg)} & \multicolumn{1}{c}{(mJy/beam)} \\
\hline
\textbf{150 MHz}\\
J0045$+$8810 &00 45 05.38 +88 10 19.19 &2019 Feb 01 & 2 & 80 &24 $\times$ 15 &53$^\circ$ & 3.5\\
J0304$+$7727 &03 04 55.06 +77 27 31.68 &2019 Jan 04 & 4 & 60 &38 $\times$ 15 &41$^\circ$ & 2.5\\
J0754$+$5324 &07 54 15.36 +53 24 56.52 &2019 Jan 04 & 4 & 60 &55 $\times$ 13 &$-$87$^\circ$ & 3.8\\
J0847$+$5723 &08 47 27.86 +57 23 35.52 &2019 Jan 04 & 4 & 60 &38 $\times$ 14 &75$^\circ$ & 3.0\\
J0858$+$7501 &08 58 33.43 +75 01 18.47 &2019 Jan 04 & 4 & 65 &47 $\times$ 14 &85$^\circ$& 3.8\\
J1326$+$5712 &13 26 50.47 +57 12 06.84 &2019 Jan 20 & 5 & 75 &26 $\times$ 18 &09$^\circ$ & 1.2\\ 
J1430$+$3649 &14 30 40.63 +36 49 07.32 &2019 Jan 20 & 5 & 75 &22 $\times$ 18 &14$^\circ$ & 1.5\\
J1536$+$8154 &15 36 59.83 +81 54 31.32 &2019 Jan 20 & 5 & 75 &32 $\times$ 17 &10$^\circ$ & 2.5\\
J1549$+$5038 &15 49 17.46 +50 38 05.32 &2019 Jan 20 & 5 & 75 &24 $\times$ 19 &09$^\circ$ & 1.5\\
J1658$+$4732 &16 58 26.54 +47 32 14.99 &2019 Feb 18 & 1 & 60 &35 $\times$ 16 &$-$78$^\circ$ & 2.5\\
J1700$+$3830 &17 00 19.96 +38 30 33.81 &2019 Feb 01 & 2 & 80 &20 $\times$ 14 &26$^\circ$ & 2.5\\
J1722$+$7046 &17 22 07.22 +70 46 28.19 &2019 Feb 01 & 3 & 90 &20 $\times$ 16 &$-$14$^\circ$ & 2.0\\
J1723$+$7653 &17 23 59.88 +76 53 11.75 &2019 Jan 20 & 5 & 75 &34 $\times$ 16 &08$^\circ$ & 2.0\\
J1846$+$4239 &18 46 42.63 +42 39 45.68 &2019 Feb 01 & 3 & 95 &21 $\times$ 16 &13$^\circ$ & 1.5\\
J2317$+$4738 &23 17 10.83 +47 38 21.48 &2019 Jan 04 & 4 & 60 &21 $\times$ 20 &23$^\circ$ & 2.5\\
\\ 
\textbf{325 MHz}\\
J0045$+$8810 &00 45 05.38 +88 10 19.19 &2019 Feb 02 & 1 & 15 &25 $\times$ 06 &69$^\circ$ & 1.0\\
J0304$+$7727 &03 04 55.06 +77 27 31.68 &2019 Jan 05 & 1 & 12 &18 $\times$ 07 &$-$01$^\circ$ & 0.6\\
J0754$+$5324 &07 54 15.36 +53 24 56.52 &2019 Jan 05 & 1 & 12 &15 $\times$ 07 &70$^\circ$ & 0.6\\
J0847$+$5723 &08 47 27.86 +57 23 35.52 &2019 Jan 05 & 1 & 12 &18 $\times$ 06 &78$^\circ$ & 1.0\\
J0858$+$7501 &08 58 33.43 +75 01 18.47 &2019 Jan 05 & 1 & 12 &22 $\times$ 06 &84$^\circ$& 0.7\\
J1326$+$5712 &13 26 50.47 +57 12 06.84 &2019 Jan 19 & 1 & 15 &12 $\times$ 07 &$-$10$^\circ$ & 0.5\\ 
J1430$+$3649 &14 30 40.63 +36 49 07.32 &2019 Jan 19 & 1 & 15 &09 $\times$ 07 &02$^\circ$ & 0.3\\
J1536$+$8154 &15 36 59.83 +81 54 31.32 &2019 Jan 19 & 1 & 15 &21 $\times$ 06 &05$^\circ$ & 0.6\\
J1549$+$5038 &15 49 17.46 +50 38 05.32 &2019 Jan 19 & 1 & 15 &11 $\times$ 07 &04$^\circ$ & 0.6\\
J1658$+$4732 &16 58 26.54 +47 32 14.99 &2019 Feb 02 & 1 & 15 &10 $\times$ 07 &09$^\circ$ & 1.0\\
J1700$+$3830 &17 00 19.96 +38 30 33.81 &2019 Feb 02 & 1 & 15 &09 $\times$ 07 &$-$12$^\circ$ & 0.5\\
J1722$+$7046 &17 22 07.22 +70 46 28.19 &2019 Feb 02 & 1 & 15 &14 $\times$ 07 &$-$22$^\circ$ & 0.8\\
J1723$+$7653 &17 23 59.88 +76 53 11.75 &2019 Jan 19 & 1 & 15 &17 $\times$ 07 &30$^\circ$ & 1.0\\
J1846$+$4239 &18 46 42.63 +42 39 45.68 &2019 Feb 02 & 1 & 15 &09 $\times$ 07 &08$^\circ$ & 0.7\\
J2317$+$4738 &23 17 10.83 +47 38 21.48 &2019 Jan 05 & 1 & 12 &13 $\times$ 08 &$-$62$^\circ$ & 0.8\\
\\
\hline
\end{tabular}
\end{table*}

\subsection{Radio Observations}
\begin{figure}
\centering	
\includegraphics[width=\columnwidth]{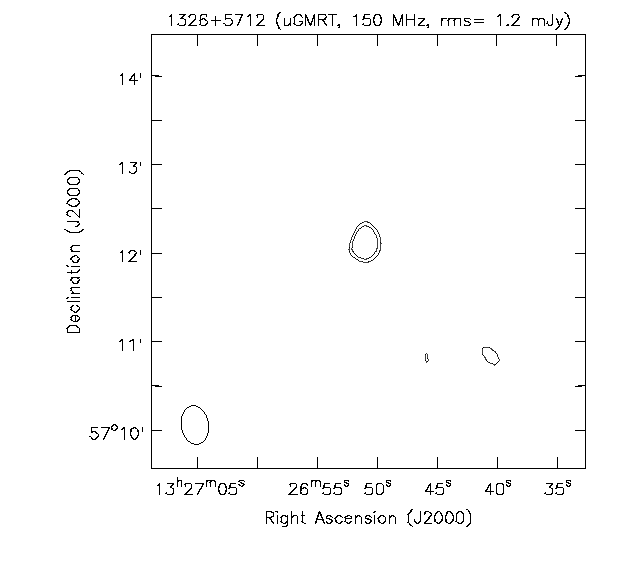} \\
\includegraphics[width=\columnwidth]{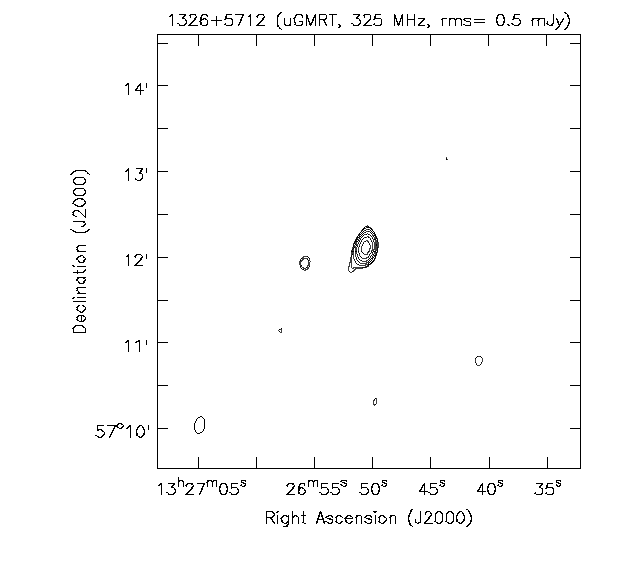}
\caption{The uGMRT contour maps of J1326$+$5712 at 150 MHz and 325 MHz, respectively. The contours are drawn at 3, 4, 8, 16, 32, 64 and 
128 times the image rms noise which is 1.2 mJy at 150 MHz and 0.5 mJy at 325 MHz. The FWHMs are 26 $\times$ 18$"$ (PA= 9$\degr$) 
and 12 $\times$ 7$"$ (PA= $-$10$\degr$) at 150 MHz and 325 MHz, respectively. The target source lies at the centre of each map.} 
\label{fig:J1326}  
\end{figure}

\begin{figure}
\centering	
\includegraphics[width=\columnwidth]{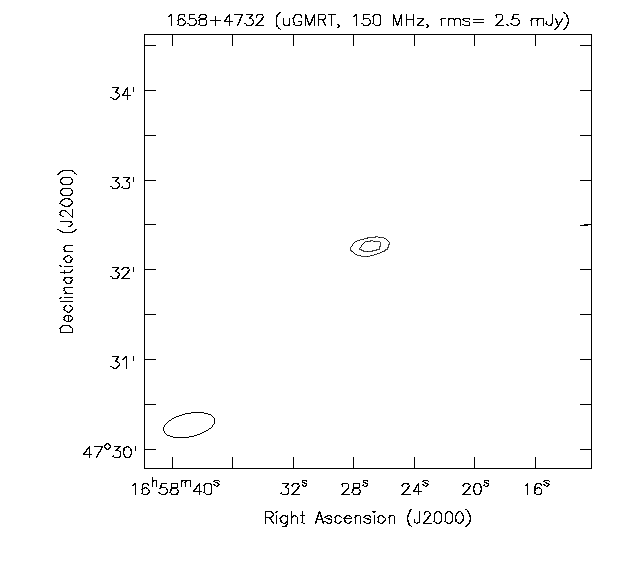} \\
\includegraphics[width=\columnwidth]{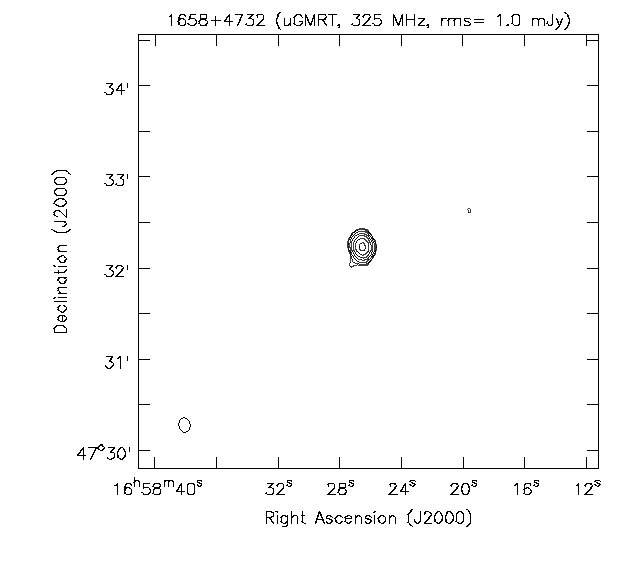} 
\caption{The uGMRT contour maps of J1658$+$4732 at 150 MHz and 325 MHz, respectively. The contours are drawn at 3, 4, 8, 16, 32, 64 and 
128 times the image rms noise which is 2.5 mJy at 150 MHz and 1.0 mJy at 325 MHz. The FWHMs are 35 $\times$ 16$"$ (PA= $-$78$\degr$) 
and 10 $\times$ 7$"$ (PA= 9$\degr$) at 150 MHz and 325 MHz, respectively. The target source lies at the centre of each map.}  
\label{fig:J1658} 
\end{figure}

The 15 EISERS candidates were observed with the recently upgraded GMRT \citep[`uGMRT',][]{Gupta2017}. The observations
were performed in a snapshot mode, quasi-simultaneously at 150 MHz and 325 MHz, between 2019 Jan 04 and 2019 Feb 18 
(Table~\ref{table:obs-log}). The integration time was 8 seconds at both 150 MHz and 325 MHz.
 One of the standard flux-density calibrators, 3C 286, 3C 48, and 3C 147, was observed at the start and
at the end of each observing session at 150 MHz. At 325 MHz, the flux-density calibrator was observed only at the start 
of the session. 
Phase calibrator(s) were observed immediately before and after each snapshot on the target source. The average total on-target time 
depended on the frequency. Since the target sources have an inverted spectrum, they are stronger at 325 MHz than at 150 MHz. 
Total observing time for each target source was about 60 minutes at 150 MHz and about 15 minutes at 325 MHz. Further 
observational details are provided in the log (Table~\ref{table:obs-log}). Flux-densities at different frequencies are listed in 
Table~\ref{table:spec-prop}, including those measured in the present uGMRT observations at 150 MHz and 325 MHz. The corresponding
radio contour maps are provided in the online material, except for the maps of the best two cases of EISERS found in this study,
namely J1326$+$5712 and J1658$+$473. These are presented in Figures~\ref{fig:J1326} \&~\ref{fig:J1658}.

\subsection{Analysis}\label{section3.2}
The measured visibilities at 150 MHz and 325 MHz were processed using the Source Peeling and Atmospheric Modelling \citep[\textsc{spam}]
[]{Intema2014} package. \textsc{spam} is a semi-automated pipeline based on \textsc{aips}, \textsc{parseltongue} and \textsc{python}. 
It performs a series of iterative flagging and calibration sequences and the imaging involves direction dependent calibration. 
This package has been used for processing of the entire TGSS data at 150 MHz \citep{Intema2017}. Details of \textsc{spam} 
and its various routines are provided in \citet{Intema2017}. 

Equation~\ref{eq:error} defines the rms uncertainty of the measured flux-density of a source, the first term is the root sum square of the
 fitting error in the \textsc{aips} task \textsc{jmfit} and the second term is the systematic error component (10\%) which includes the error arising from the elevation dependent gains of the antennas \citep{Chandra2004}. 
\begin{equation} \label{eq:error}
\sqrt{ \rm{(map~rms)}^{2} + (10 \%~\rm{of~the~peak~flux})^{2}}
\end{equation}

\begin{landscape}
\begin{figure}
\vspace{-2.0cm}
\hspace{-2.3cm}
\includegraphics[width=1.6\textwidth]{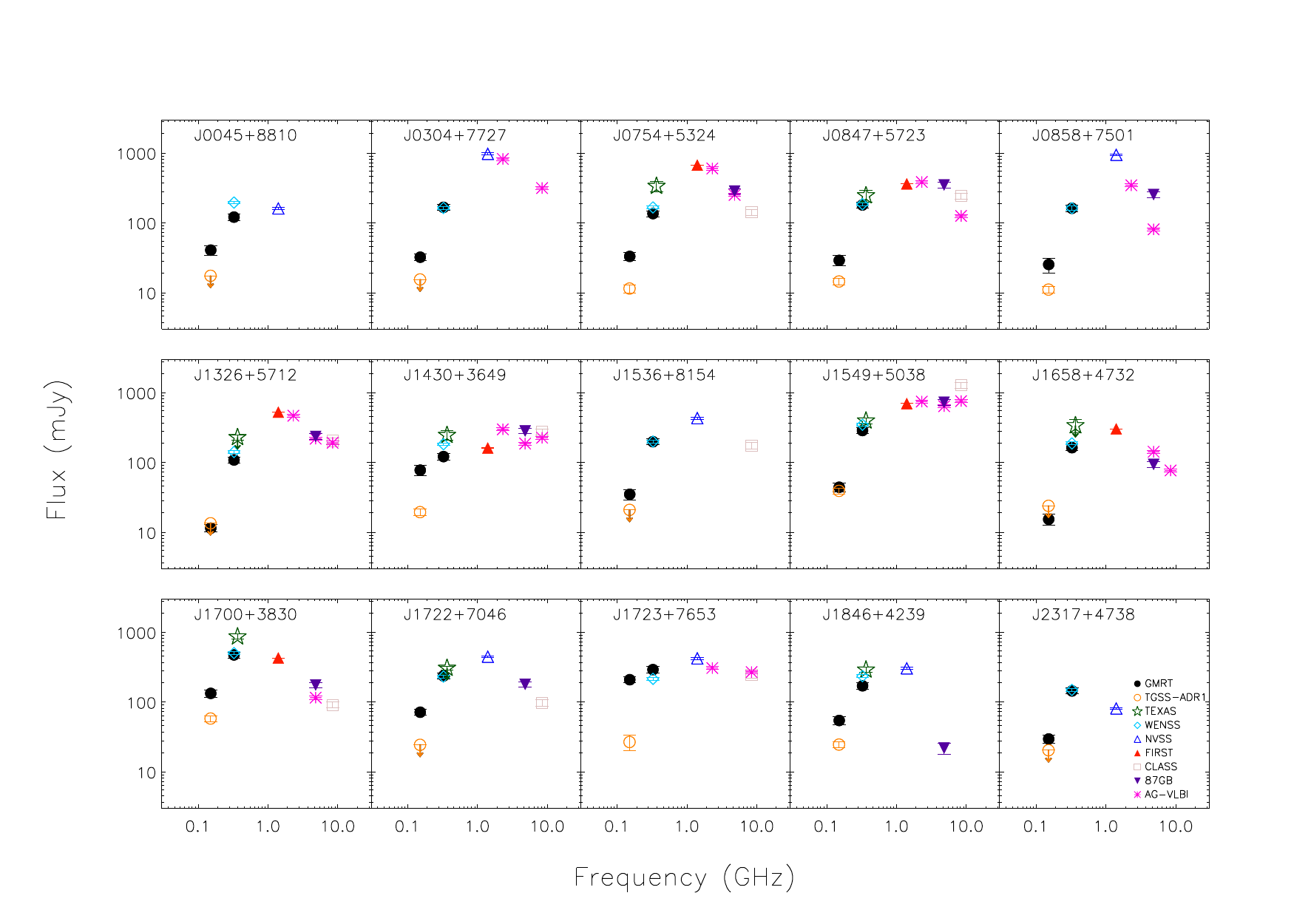}
\vspace{-1.0cm}
\caption{Radio spectra of the 15 EISERS candidates (see Table ~\ref{table:spec-prop}). In the last panel, the names of the surveys 
are listed adjacent to the corresponding symbols used (the frequency of the survey increases downwards). The filled circles represent 
the measurements from the present uGMRT observations at 150 MHz and 325 MHz.} \label{fig:spec_all} 
\end{figure}
\end{landscape}

In Table~\ref{table:spec-prop} the GMRT flux densities found here at 150 and 325 MHz are listed together with those at other frequencies, taken from the literature. Based on these, the radio spectra of the 15 sources are displayed in 
Figure~\ref{fig:spec_all}.

Bearing in mind the possibility of a substantial bias in the flux-density scales of the present uGMRT maps at 150 MHz, we
have determined flux scaling factors (FSF) by comparing these uGMRT maps with their counterparts in the TGSS-ADR1 survey
\citep{Intema2017}, also at 150 MHz. For a given target field, FSF was determined by taking an average of the ratios 
of the flux densities of several relatively bright discrete sources seen in both the maps. Such sources were selected within 
1$^{\circ}$ of the target source in its uGMRT map at 150 MHz, taking care that they are located in relatively isolated (i.e., unconfused) 
parts of the map and, moreover, are stronger than $\sim$75 mJy at 150 MHz. The FSF values for all the 15 target fields
are given in Table~\ref{table:FSF}. Thus, for any source seen in the 150 MHz uGMRT map of a given target field, multiplying
its flux density with the estimated FSF for that field, would translate its uGMRT flux-density to the flux-density scale of the TGSS-ADR1 survey. The correspondingly adjusted values of spectral index $\alpha$ (150-325 MHz) of the observed EISERS candidates
are listed in the last column of Table~\ref{table:FSF}. Note that this is just a cautionary step, meant to provide an additional check 
on the reliability of our estimated spectral indices of the EISERS candidates, and this procedure is not intended for application to
uGMRT image analyses, in general. The small number of the fields for which FSF values have been determined, does not allow us to make a general comment about the origin of the occasionally significant departures of FSF values from unity.

\begin{landscape}
\begin{table}
\addtolength{\tabcolsep}{-4pt}
\caption{Optical identifications, redshifts and radio flux densities of the 15 EISERS candidates.}
\label{table:spec-prop}
\vspace{-0.5cm}
\begin{tabular}{ccccccccccccc}\\
\hline

\multicolumn{1}{c}{Source}& \multicolumn{1}{c}{Optical$^{a}$}  & \multicolumn{1}{c}{150 MHz} & \multicolumn{1}{c}{325 MHz} & \multicolumn{1}{c}{150 MHz} & \multicolumn{1}{c}{325 MHz}& \multicolumn{1}{c}{365 MHz}& \multicolumn{1}{c}{1.4 GHz}&\multicolumn{1}{c}{2.3 GHz}&\multicolumn{1}{c}{4.8 GHz} &\multicolumn{1}{c}{4.8 GHz} & \multicolumn{1}{c}{8.4 GHz}&\multicolumn{1}{c}{8.4 GHz}  \\
(J2000)& ID & uGMRT$^{b}$ & uGMRT$^{b}$ & TGSS-ADR1$^{b}$ &WENSS$^{b}$&TEXAS$^{b}$ & FIRST$^{b}$ & AG-VLBI$^{b}$ &87GB$^{b}$& AG-VLBI$^{b}$ & CLASS$^{b}$&AG-VLBI$^{b}$  \\
  & (z)    & (mJy) & (mJy) &(mJy) &(mJy) &(mJy)& (mJy) &(mJy) &(mJy)& (mJy) &(mJy)  &(mJy)  \\
\hline
0045+8810& & 40.8${\pm}$6.6 &121.2${\pm}$12.2 &$<$17.5	    & 196.0${\pm}$08.3&  &161.1${\pm}$04.8$^{c}$& &		  & & &\\

0304+7727& &32.2${\pm}$3.7 &167.2${\pm}$16.7&$<$15.5       & 164.0${\pm}$07.2& & 976.3${\pm}$29.3$^{c}$& 825${\pm}$41& & & & 316${\pm}$16  \\

0754+5324& &33.2${\pm}$4.5 &136.2${\pm}$13.6&11.5${\pm}$1.2& 167.0${\pm}$09.0& 337${\pm}$55& 676.2${\pm}$2.6 & 603${\pm}$30 &285.0${\pm}$25.0& 252${\pm}$13 &141.7${\pm}$14.1 & \\

0847+5723 & &29.1${\pm}$4.8 &180.9${\pm}$18.1 &14.5${\pm}$1.5& 186.0${\pm}$10.0&247${\pm}$49 &363.2${\pm}$0.14 &386${\pm}$19 & 350.0${\pm}$31.0&  &242.6${\pm}$24.2& 126${\pm}$06 \\

0858+7501 & &25.4${\pm}$6.1 &162.2${\pm}$16.2 &11.1${\pm}$1.1& 163.0${\pm}$12.6& &947.4${\pm}$28.4$^{c}$& 346${\pm}$17 &255.0${\pm}$23.0& 81${\pm}$04 &		 &  \\

1326+5712 & &11.5${\pm}$1.4 &108.9${\pm}$10.9 &$<$13.5       & 142.0${\pm}$07.4& 230${\pm}$20 & 528.5${\pm}$0.16 & 469${\pm}$23 &237.0${\pm}$21.0& 220${\pm}$11 &204.8${\pm}$20.4& 192${\pm}$10 \\

1430+3649 & Q(z=0.560) &77.8${\pm}$13.4 &121.8${\pm}$12.2 &19.5${\pm}$2.0& 183.0${\pm}$08.6& 247${\pm}$49 &162.0${\pm}$0.15& 299${\pm}$15 &283.0${\pm}$25.0& 187${\pm}$09  &273.5${\pm}$27.3& 227${\pm}$11 \\

1536+8154& &35.1${\pm}$5.9 & 198.9${\pm}$19.9 &$<$21.0       & 200.0${\pm}$08.5& &432.3${\pm}$13.0$^{c}$&  &  & &174.9${\pm}$17.4 & \\

1549+5038 & Q(z=2.174) &44.3${\pm}$6.2 & 287.8${\pm}$28.8 &39.3${\pm}$3.9& 348.0${\pm}$14.6&397${\pm}$35& 701.4${\pm}$0.14&748${\pm}$37 &731.0${\pm}$65.0& 642${\pm}$32 & 1286.1${\pm}$128.6& 760${\pm}$38 \\

1658+4732 & &15.4${\pm}$2.8 & 163.4${\pm}$16.4 &$<$24.0       & 188.0${\pm}$09.3& 340${\pm}$72&304.6${\pm}$0.18 &  & 094.0${\pm}$10.0&	&	&  \\

1700+3830& &133.7${\pm}$17.2 & 475.5${\pm}$47.6 &58.3${\pm}$5.8& 506.0${\pm}$20.9&868${\pm}$74& 430.2${\pm}$0.14 &  & 176.0${\pm}$16.0& 116${\pm}$06  &091.3${\pm}$9.1& \\

1722+7046 & &71.8${\pm}$7.2 & 238.1${\pm}$23.8 &$<$24.5       & 233.0${\pm}$11.2& 308${\pm}$17 &447.8${\pm}$13.4$^{c}$&  & 180.0${\pm}$16.0&  & 098.2${\pm}$9.8 & \\

1723+7653& Q (z=0.680) &209.9${\pm}$21.1 & 292.8${\pm}$29.5 &26.9${\pm}$6.9 & 216.0${\pm}$09.1& &423.3${\pm}$12.7$^{c}$& 309${\pm}$15 & &	& 244.1${\pm}$24.4& 270${\pm}$14 \\

1846+4239& &54.7${\pm}$7.2 & 171.6${\pm}$17.2 &24.7${\pm}$2.5& 236.0${\pm}$10.4& 291${\pm}$26 &305.3${\pm}$09.2$^{\ast}$& &   22.0${\pm}$04.0&	& 	&   \\

2317+4738 & &29.7${\pm}$3.7 & 145.0${\pm}$14.5 &$<$20.5       & 150.0${\pm}$07.9& &081.3${\pm}$02.5$^{c}$& & &	&	  	&	   \\
\hline
\end{tabular}

{$^{a}$ Q -- Quasar (\citet{Paris2017}, \citet{Veron2010}). $^{b}$ TGSS-ADR1--\citet{Intema2017}; WENSS -- \citet{deBruyn2000}; TEXAS -- \citet{Douglas1996}; FIRST -- \citet{Helfand2015}; AG-VLBI -- Astrogeo VLBI FITS Image Database; 87GB -- \citet{Gregory1991}; CLASS -- \citet{Myers2003}; NED -- NASA Extragalactic Database.}\\
{$^{c}$ Outside the FIRST Survey range, hence the flux density is taken from the NVSS \citep{Condon1998}.}\\
\end{table}
\end{landscape}
\clearpage

\clearpage

\begin{table}
\addtolength{\tabcolsep}{-5pt}
\caption{Spectral Index, $\alpha_{thick}$ (150-325 MHz, uGMRT),  $\alpha_{thin}$ (1400-4850 MHz) and the final status of the 15 EISERS candidates.}
\label{table:summary}
\vspace{-0.5cm}
\begin{tabular}{ccccc}\\
\hline

\multicolumn{1}{c}{Source} & \multicolumn{1}{c}{Spectral Index}  & \multicolumn{1}{c}{Spectral Index} & \multicolumn{1}{c}{log P$_{5 GHz}$}& \multicolumn{1}{c}{EISERS}$^{a}$ \\
 & $\alpha_{thick}$ & $\alpha_{thin}$& W Hz$^{-1}$ & Status\\
  & (150-325 MHz) & (1400-4850 MHz) & & \\
\hline
0045+8810 & 1.41${\pm}$0.25 &$-$ & $-$   &N\\
0304+7727 & 2.13${\pm}$0.20 & -0.63${\pm}$0.03$^{b}$& 27.24$^{c}$   &N\\
0754+5324 & 1.83${\pm}$0.22 & -0.70${\pm}$0.07& 27.02$^{c}$&N\\
\textbf{0847+5723} & 2.36${\pm}$0.25 & -0.03${\pm}$0.07& 27.11$^{c}$&T\\
\textbf{0858+7501} & 2.40${\pm}$0.34 & -1.06${\pm}$0.08& 26.97$^{c}$&T\\
\textbf{1326+5712} & 2.91${\pm}$0.20 & -0.65${\pm}$0.07& 26.94$^{c}$&Y\\
1430+3649 & 0.58${\pm}$0.26 & 0.45${\pm}$0.07& 26.43      &N\\
1536+8154 & 2.24${\pm}$0.25 & -0.50${\pm}$0.06$^{b}$& 26.93$^{c}$   &N\\
\textbf{1549+5038} & 2.42${\pm}$0.22 & 0.03${\pm}$0.07& 28.26      &T\\
\textbf{1658+4732} & 3.05${\pm}$0.27 & -0.95${\pm}$0.09& 26.54$^{c}$&Y\\
1700+3830 & 1.64${\pm}$0.21 & -0.72${\pm}$0.07& 26.81$^{c}$&N\\
1722+7046 & 1.55${\pm}$0.22 & -0.73${\pm}$0.08& 26.82$^{c}$&N\\
1723+7653 & 0.43${\pm}$0.18 & -0.31${\pm}$0.06$^{b}$& 26.65        &N\\
1846+4239 & 1.48${\pm}$0.21 & -2.12${\pm}$0.15& 25.91$^{c}$&N\\
2317+4738 & 2.05${\pm}$0.21 &$-$ & $-$   &N\\
\hline
\end{tabular}

{$^{a}$ N- Not an EISERS, T- Tentative EISERS, Y- Bonafide EISERS}.\\ 
{$^{b}$ $\alpha_{thin}$ (1400-8400 MHz)}\\
{$^{c}$ z=1 is assumed to calculate the radio power wherever the redshift information is not available.}\\
{The bonafide and tentative EISERS are marked in bold.}\\ 
\end{table}

\begin{table}
\small
\centering
\caption{Flux-density Scaling Factors (FSFs) and the correspondingly adjusted values of $\alpha$ (150$-$325 MHz).} 
\label{table:FSF}
\begin{tabular}{ccc}\\
\hline

\multicolumn{1}{c}{Source} & \multicolumn{1}{c}{FSF} & \multicolumn{1}{c}{$\alpha_{TGSS}^{a}$}  \\
\multicolumn{1}{c}{} & \multicolumn{1}{c}{TGSS/uGMRT}& \multicolumn{1}{c}{} \\
\hline
J0045$+$8810 & 1.185&1.19${\pm}$0.23\\
J0304$+$7727 & 0.810&2.40${\pm}$0.19\\
J0754$+$5324 & 1.139&1.66${\pm}$0.25\\
\textbf{J0847$+$5723} & 1.010&2.35${\pm}$0.23\\
\textbf{J0858$+$7501} & 0.946&2.47${\pm}$0.26\\
\textbf{J1326$+$5712} & 0.771&3.24${\pm}$0.20\\ 
J1430$+$3649 & 0.597&1.25${\pm}$0.15\\
J1536$+$8154 & 1.012&2.23${\pm}$0.21\\
\textbf{J1549$+$5038} & 0.863&2.61${\pm}$0.18\\
\textbf{J1658$+$4732} & 1.577&2.47${\pm}$0.32\\
J1700$+$3830 & 0.427&2.74${\pm}$0.14\\
J1722$+$7046 & 1.089&1.44${\pm}$0.19\\
J1723$+$7653 & 0.390&1.65${\pm}$0.14\\
J1846$+$4239 & 0.679&1.98${\pm}$0.16\\
J2317$+$4738 & 0.688&2.53${\pm}$0.19\\
\hline
\end{tabular}

{$^{a}$ Estimate of spectral Index (150-325 MHz) when FSF has been applied in order to scale the present 150 MHz flux 
density to the flux scale of the TGSS-ADR1 survey. The uGMRT flux value at 325 MHz along with the rescaled value at 150 MHz are used for calculating $\alpha_{TGSS}$ (Section \ref{section3.2}).}\\
{The bonafide and tentative EISERS are marked in bold.}\\ 
\end{table}

\section{\textbf{Comments on Individual Sources}}\label{section4}
\subsection*{\textbf{J0045$+$8810}}
This source is not seen in the TGSS-ADR1 map at 150 MHz, albeit the higher sensitivity of the present uGMRT map ($\sigma$ $=$ 3.5 
mJy, Figure 1: online material) has led to its detection at this frequency. At 325 MHz, while the source is clearly detected in the 
present map, its flux density is much lower ($\sim$40\%) than the value we had initially taken from the WENSS catalogue. Accordingly,
the present estimate of its $\alpha$ (150-325 MHz) = $+$1.41 ${\pm}$ 0.25 (Table~\ref{table:spec-prop}), which is not large enough to 
qualify it as an EISERS.

\subsection*{\textbf{J0304$+$7727}}
This EISERS candidate, although not seen in the TGSS-ADR1 map, is clearly present in our uGMRT map at 150 MHz ($\sigma$ $=$ 2.5 mJy, 
Figure 2: online material). At 325 MHz, its uGMRT flux density matches well with that given in the WENSS catalogue. With the presently
measured spectral index $\alpha$ $=$ $+$2.13${\pm}$0.20, the source does not qualify as an EISERS. Its VLBA image at 8.6 GHz shows a 
pair of unresolved components separated by $\sim$10 mas \citep{Sokolovsky2011}. Although the available spectral measurements fall 
short of pinning down its spectral turnover frequency, it is unlikely to be below 0.5 GHz (Figure~\ref{fig:spec_all}). 

\subsection*{\textbf{J0754$+$5324}}
The source is clearly seen in the present uGMRT map at 150 MHz ($\sigma \sim$ 3.8 mJy, Figure 3-online material), at a flux density 
level consistent with the TGSS-ADR1 value (Table~\ref{table:spec-prop}). At 325 MHz, its flux density found here is $\sim$
20\% lower than that in the WENSS catalogue. Its spectral index $\alpha$ = $+$1.83${\pm}$0.22 does not qualify it as an EISERS (Table~\ref{table:spec-prop}). Its GPS\footnote{GPS sources are defined as having an integrated spectrum that shows a single peak near 1 GHz and steep slopes on either side of the peak, cf. \citep{Spoelstra1985, GopalKrishna1983, GopalKrishna1993, deVries1997, Odea1998, An2012}} type radio spectrum (Figure~\ref{fig:spec_all}) indicates that it could be a `compact symmetric object' (CSO) \citep{Phillips1982, Conway1994, Wilkinson1994}. This is indeed borne out by its VLBA image at 5 GHz which shows two components separated by $\sim$9 mas \citep{Helmboldt2007}. An earlier VLBI image at 8.4 GHz had shown this source to be dominated by two hot spots with a bright jet pointing toward the northernmost lobe \citep{Peck2000}. The `core' appears to be a knot in the jet, moving toward the northwestern hot spot at 0.060${\pm}$0.026 mas per year \citep{Gugliucci2005}. 

\subsection*{\textbf{J0847$+$5723}}
The source is clearly detected in the present uGMRT map at 150 MHz ($\sigma \sim$ 3.0 mJy, Figure 4-online material), at a 
flux-density level $\sim$50\% higher than that in the TGSS-ADR1 map. Note that we find the flux scales of these two 150 MHz maps 
to be in good agreement (Table~\ref{table:FSF}). At 325 MHz, its flux-density measured from our uGMRT map is consistent with the 
WENSS catalogue value. With an $\alpha$ (150 - 325 MHz) = $+$2.36${\pm}$0.25, this source is very close to the SSA limit of $+$2.5. 
\citet{Helmboldt2007} have placed it in the \lq long-jet\rq~morphology class, on the basis of its VLBI image at 5 GHz.  
(Figure~\ref{fig:spec_all}). The source is also detected at 15 GHz \citep[66.9${\pm}$6.7 mJy,][]{Taylor2005} and 30 GHz \citep[43${\pm}$4 mJy,][]{Lowe2007}, suggesting a compact component becoming dominant at these high frequencies (Figure~\ref{fig:spec_all}). Note also that a flux variation of $\sim$50\% has been reported at 8.4 GHz, based on the VLBI images taken almost two decades apart (Table~\ref{table:spec-prop}).  

\subsection*{\textbf{J0858$+$7501}}
There is a good agreement between its flux densities measured on the TGSS-ADR1 and the present uGMRT maps at 150 MHz. The same 
holds for its 325 MHz flux densities measured from the present uGMRT and the WENSS maps. Its spectral slope $\alpha$ = $+$2.40${\pm}$0.34 
(Table~\ref{table:spec-prop}) closely approaches the SSA limit of $+$2.5. Note that this source was also included in the List 3 
of GPS sources \citep{GopalKrishna1993}. There is a large difference between its 4.8 GHz flux density (Table~\ref{table:spec-prop}) 
taken from the 87GB catalogue and that measured from the high-resolution (VLBI) map presented in the AstroGeo VLBI calibrator 
catalogue (henceforth, AG-VLBI) \footnote{http://astrogeo.org/vlbi$\_$images/}. 

\subsection*{\textbf{J1326$+$5712}}
Although this source has been detected in the present uGMRT observations at a low flux level of 11.5${\pm}$1.4 mJy at 150 MHz, it is 
expectedly not seen in the TGSS-ADR1 image at 150 MHz. At 325 MHz, its uGMRT flux-density reported here is only $\sim$75\% of the 
WENSS value. Still, with a spectral index $\alpha$ (150 - 325 MHz) $=$ $+$2.91${\pm}$0.20 (Table~\ref{table:spec-prop}), this source 
is probably the most secure example of an EISERS. Its present uGMRT maps at 150 MHz and 325 MHz are shown in Figure~\ref{fig:J1326}. 
In the AG-VLBI calibrator catalogue, it is seen as a double source, both at 5 and 8.4 GHz, hence consistent with a CSO classification. In the Effelsberg observations at 10.5 GHz, the source was found to be unpolarised to a limit of 2\% \citep{Pasetto2016}. 

\subsection*{\textbf{J1430$+$3649}}
The presently measured flux-density of 77.8${\pm}$13.4 mJy at 150 MHz is approximately four times the value reported in the 
TGSS-ADR1 (Table~\ref{table:spec-prop}). Such a massive flux variability at metre wavelengths is highly unlikely. Note that the 
flux scales of the two maps are also found to differ by a large factor of $\sim$2 (Table~\ref{table:FSF}). Thus, based on the 
present measurement at 150 MHz, this source can be clearly discounted as an EISERS (Table~\ref{table:spec-prop}). The spectrum 
remains mildly inverted at least up to 10 GHz (Figure~\ref{fig:spec_all}). The source is identified with a quasar located at 
$z$ = 0.5660 \citep{Paris2017}. Its VLBA observations at 5 GHz reveal a single source, with a mean radius of 
1.3 mas \citep{Helmboldt2007}.

\subsection*{\textbf{J1536$+$8154}}
Absent in the 150 MHz TGSS-ADR1 map, this source is detected at $\sim$10$\sigma$ level in the present uGMRT map, with a 
flux-density of 35.1 mJy at 150 MHz. Note that the flux scales of these two maps are in excellent agreement 
(Table~\ref{table:FSF}). At 325 MHz, its uGMRT flux-density agrees well with the WENSS value (Table~\ref{table:spec-prop}). With a spectral index $\alpha$ (150 - 325 MHz) = $+$2.24${\pm}$0.25 (Table~\ref{table:spec-prop}), the source is not confirmed as an EISERS. 

\subsection*{\textbf{J1549$+$5038}}
Its present uGMRT flux density at 150 MHz is in excellent agreement with the TGSS-ADR1 value. At 325 MHz, we find it to be 
$\sim$20\% weaker compared to the WENSS flux density. With our estimated $\alpha$ (150 - 325 MHz) of $+$2.42${\pm}$0.22, the 
source is a borderline EISERS and hence worthy of follow-up. 
In the VLBA images at 5 GHz \citep{Xu1995,Helmboldt2007} and at 2 and 8 GHz \citep{Fey2000} the source appears resolved into 
a dominant flat-spectrum core and a fairly bright curved jet which is $\sim$10 mas long and extends towards the South-West. The jet 
is itself resolved into a couple of knots. The source is also detected at 15 GHz \citep[810${\pm}$03 mJy,][]{Richards2011}, as well as
at 30 GHz (738${\pm}$39 mJy). Moderately strong flux variability, with a modulation index of 9\%, has been reported at 15 GHz in 
\citet{Richards2014}. \citet{Pasetto2016} have inferred a large (rest-frame) rotation measure (RM = 1400$\pm$500 rad $m^{-2}$), 
revealing the presence of dense magneto-ionic plasma. In the SDSS quasar catalogue-DR14, this source has been identified with a 
quasar at $z$ = 2.174 \citep{Paris2017}. The data provided in Table~\ref{table:spec-prop} and Figure~\ref{fig:spec_all} confirm
its strong flux variability at centimetre wavelengths.

\subsection*{\textbf{J1658$+$4732}}
Although not seen in the TGSS-ADR1 map at 150 MHz, this source is detected in the present uGMRT image, with a flux 
density of 15.4${\pm}$2.8 mJy at 150 MHz (Figure~\ref{fig:J1658}). At 325 MHz, its present measurement of flux-density is 
marginally ($\sim$10\%)
lower than that reported in the WENSS catalogue (Table~\ref{table:spec-prop}). With a spectral index $\alpha$ (150 - 325 MHz) 
= $+$3.05${\pm}$0.27, this source is securely confirmed as a bona fide EISERS. Its AG-VLBI image at 4.3 GHz, shows a core with a jet extending towards the
South-East, possibly terminating in a lobe $\sim$60 mas away from the core.

\subsection*{\textbf{J1700$+$3830}}
In the present uGMRT observations at 150 MHz, this source appears at $\sim$ 2 times stronger level than that seen in the 150 MHz 
TGSS-ADR1 image (Table~\ref{table:spec-prop}). Hence, with a revised spectral index of $\alpha$ (150 - 325 MHz) = $+$1.64${\pm}$0.21, it
is clearly discounted as an EISERS. Its VLBA image at 5 GHz, made under the VIP survey \citep{Helmboldt2007}, shows it to be a 
core with a jet extending eastward, possibly terminating in a lobe $\sim$50 mas away. The source has a GPS type spectrum peaking 
near 0.5 GHz (Figure~\ref{fig:spec_all}).

\subsection*{\textbf{J1722$+$7046}}
This source, not seen at 150 MHz in the TGSS-ADR1 map, is clearly detected in the present uGMRT observations, at a high flux 
level of 71.8${\pm}$7.2 mJy at 150 MHz. Hence, with a revised spectral index of $\alpha$ (150-325 MHz) = $+$1.55${\pm}$0.22 it 
does not qualify as an EISERS. 

\subsection*{\textbf{J1723$+$7653}}
Barely detected in the TGSS-ADR1 map at 150 MHz (26.9${\pm}$6.9 mJy, Table~\ref{table:spec-prop}), the source is detected in the present uGMRT observations, at a very strong level (209.9${\pm}$21.1 mJy, Table~\ref{table:spec-prop}). We find that another few fairly bright sources seen in our 150 MHz uGMRT map are absent in the 150 MHz TGSS-ADR1 map. The consequent drastic revision of its spectral index to $\alpha$ (150-325 MHz) = +0.43${\pm}$0.18 leaves this source far removed from the threshold of $\alpha$ (150-325 MHz) = +2.5 defined for EISERS. The source is identified as a quasar at $z$ = 0.680 \citep{Veron2010}. Its AG-VLBI image at 8.4 GHz, shows a core with a 5 mas long jet-like eastward protrusion. 

In our sample, this is the only source to show a clear indication of being resolved in the present uGMRT observations at
325 MHz. The dominant central component $\sim$285 mJy at 325 MHz) is straddled by two lobes separated by $\sim$40 arcsec along position angle $\sim$120$^{\circ}$ (Figure 11, online material). Flux densities of the northeastern and the southwestern lobes are roughly 20 and 5 mJy at 325 MHz. The flux of this source at 151 MHz in the 6C survey of radio sources \citep{Hales1991} is 300 mJy. This is more than 10 times the flux reported in TGSS-ADR1. Thus, its integrated spectrum remains nearly flat from about 0.1 to 10 GHz (Figure~\ref{fig:spec_all}). 

\subsection*{\textbf{J1846$+$4239}}
Its flux density measured on the present 150 MHz uGMRT map is nearly twice the value given in the 150 MHz TGSS-ADR1 catalogue. 
Thus, the present estimate of its spectral index $\alpha$ (150 - 325 MHz) = $+$1.48${\pm}$0.21 clearly disqualifies it as an 
EISERS (Table~\ref{table:spec-prop}).

\subsection*{\textbf{J2317$+$4738}}
This EISERS candidate, undetected in the 150 MHz TGSS-ADR1 map is clearly visible in the present uGMRT map, with a flux density 
of 29.7${\pm}$3.7 mJy at 150 MHz (Table~\ref{table:spec-prop}). At 325 MHz, its uGMRT flux of 145.0${\pm}$14.5 mJy is consistent 
with the WENSS value (Table~\ref{table:spec-prop}). With a spectral index of $\alpha$ (150 - 325 MHz) $=$ $+$2.05${\pm}$0.21, 
this source fails to qualify as an EISERS.  

\section{Spectral Modelling}\label{section5}
As discussed in Paper I and Paper II, an inverted radio spectrum with a slope $\alpha$ $>$ $\alpha_c$ $=$ $+$2.5 would 
be inconsistent with the standard SSA interpretation of the spectral turnover in extragalactic radio sources, within the framework 
of the canonical (i.e., power-law) energy distribution of the radiating relativistic electrons \citep{Slish1963, Kellermann1969, 
Pacholczyk1970, Rybicki1979}. As an alternative explanation, free-free absorption (FFA) has often been invoked \citep{Kellermann1966, 
Bicknell1997, Kuncic1998, Kameno2000, Vermeulen2003, Stawarz2008, Callingham2015}. VLBI observations have already furnished substantial evidence for FFA effects occurring in extragalactic radio sources, albeit only for the nuclear region. Prominent 
examples of flux attenuation on parsec scale due to FFA include 3C 345 \citep{Matveenko1990}, Centaurus A \citep{Jones1996, 
Tingay2001}, Cygnus A \citep{Krichbaum1998} and 3C 84 \citep{Walker1994, Levinson1995}. Whether FFA can attenuate practically the 
entire radio emission of a source, as inferred from the sharp turnover of some of the integrated radio spectra displayed in 
Figure~\ref{fig:spec_all}, remains to be demonstrated. 

In the following subsections, we shall briefly examine the two absorption models, followed by their application to the spectral measurements for the present sample of sources.

\subsection{Synchrotron self-absorption (SSA)}
 SSA, a frequently invoked mechanism to explain the low-frequency absorption of radio spectra, was first discussed by \citet{Slish1963} and 
\citet{Kellermann1966} in the context of GPS type spectra.  
A synchrotron source cannot have a brightness temperature exceeding the plasma temperature of the radiating non-thermal electrons. This leads to a spectral turnover such that the spectral slope at frequencies below the turnover cannot exceed a value of $+$2.5. It is valid for any slope of the energy spectrum of the radiating particles having a \lq\lq power-law\rq\rq \citep{Slish1963}. Thus, in the high opacity limit

\begin{equation} \label{ssa1}
S_{\nu} \propto \nu^{5/2}
\end{equation}

\citet{Tingay2003} have parameterised the spectrum in terms of the power law index, $\beta$, of the electron energy distribution, 
such that $\alpha$= $-$($\beta$ $-$ 1)/2. Further, assuming that the synchrotron source is homogeneous, the spectrum is modelled 
by, 

\begin{equation} \label{ssa2}
S_{\nu} = a_1 \left(\frac{\nu}{\nu_{p}}\right)^{-\left(\beta-1\right)/2}     \left(\frac{1-e^{-\tau}}{\tau}\right)
\end{equation}
\\
where,
\begin{equation} \label{ssa3}
\tau= \left(\frac{\nu}{\nu_{p}}\right)^{-\left(\beta+4\right)/2}
\end{equation}

$a_1$ is the normalization parameter of the intrinsic synchrotron spectrum and $\nu_{p}$ corresponds to the frequency at which the source becomes optically thick. 

\subsection{Free-free absorption (FFA)}
FFA is the absorption of the synchrotron radiation (from relativistic electrons) by a distribution of thermal plasma, either external 
or internal to the synchrotron emitting volume. In each of the cases, the thermal plasma could be inhomogeneous or essentially 
homogeneous. Such possibilities have been considered in several studies \citep[e.g.][]{Bicknell1997, Tingay2003, Callingham2015}.
Below, we reproduce the following cases:

Case 1: Assuming that a homogeneous ionized screen surrounds a standard synchrotron radio source, the free-free absorbed spectrum 
is parameterised by,

\begin{equation} \label{ffa3}
S_{\nu} = a_2 \nu^{\alpha} e^{-\tau}
\end{equation}

where \lq $a_2$\rq~is the normalization parameter of the intrinsic synchrotron spectrum, $\alpha$ is its spectral index. The optical depth is parameterised by $\tau$ = $(\nu/\nu_{p})^{-2.1}$, where $\nu_{p}$ is the frequency at which the optical depth becomes unity \citep{Kellermann1966, Tingay2003}.
\\
\\
Case 2: Here the absorbing ionised plasma is located inside the synchrotron source, hence it is the case of internal FFA \citep{Kellermann1966,Tingay2003, Callingham2015} for which 

\begin{equation} \label{ffa4}
S_{\nu} = a_3 \nu^{\alpha} \left(\frac{1 - e^{-\tau}}{\tau_{\nu}}\right)
\end{equation}
\\
\\
Case 3: \citet{Bicknell1997} introduced an alternative FFA model wherein the FFA screen is inhomogeneous and external to the synchrotron emitting lobes of a double radio source. In this model, a bow shock produced by the advancing jet creates an external, inhomogeneous FFA screen as it photo-ionises the thermal gas clouds filling the ambient (interstellar) medium. Expectedly, the ionised gas clouds would have a range of optical depth which \citet{Bicknell1997} assume to follow a power-law distribution $\eta$, parameterized by the index p, such that $\eta$ $\propto$ $\int{}{} (n_{e}^2T_{e}^{-1.35})^p dl$. The model can be represented by,

\begin{equation} \label{ffa5}
S_{\nu} = a_4\left(p+1\right)\gamma \left\lbrack p+1, \left(\frac{\nu}{\nu_{p}}\right)^{-2.1}\right\rbrack \left(\frac{\nu}{\nu_{p}}\right)^{2.1\left(p+1\right)+\alpha}
\end{equation}

where the term $\gamma \left[p+1, .. \right]$ represents the incomplete gamma function.

From their recent simulations of relativistic jets interacting with a warm, inhomogeneous medium, \citet{Bicknell2018} have argued that free-free absorption can account for the spectral peaking at $\sim$GHz frequencies and the power-law shaped inverted radio spectrum as revealed by the radio observations. A more physically motivated scenario involving log-normal opacity distribution substitutes for the power-law parametrisation of the free-free optical depth (see above). The situation envisaged in this model is particularly relevant to young radio sources. Unfortunately, spectral fitting cannot be done for this case, due to the non-availability of an analytical spectral model corresponding to their simulations.

\subsection{Fitting}\label{section5.3}
For all the EISERS candidates, we have got spectral data in the range from 150 MHz to 8 GHz. We model the spectrum of a given source by fitting various absorption scenarios represented by equations 3, 5, 6 and 7. The best-fit models are obtained by comparing the reduced $\chi^{2}$ values for the fits and these are presented in Table~\ref{table:fitting}.

\begin{figure*}
\centering	
\includegraphics[width=1.5\columnwidth]{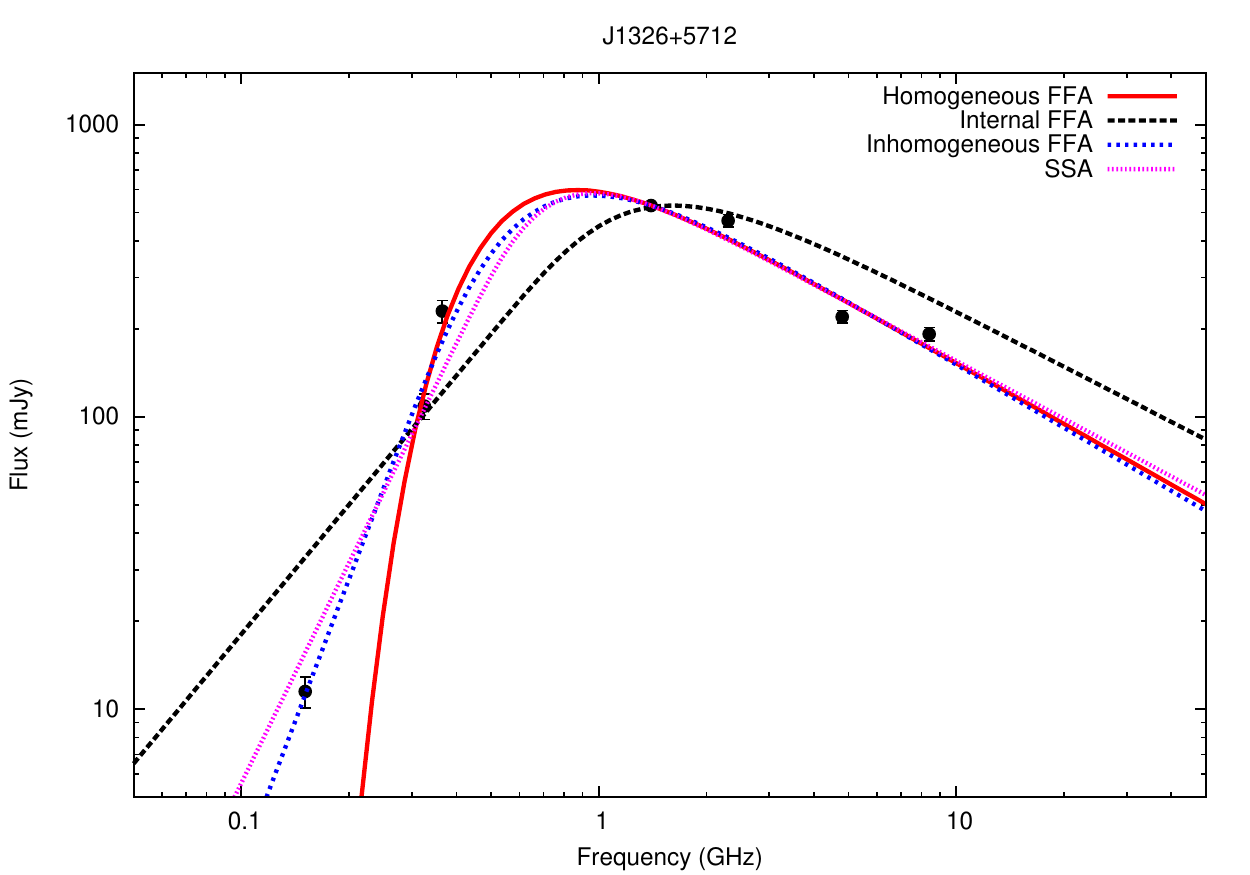} \\
\includegraphics[width=1.5\columnwidth]{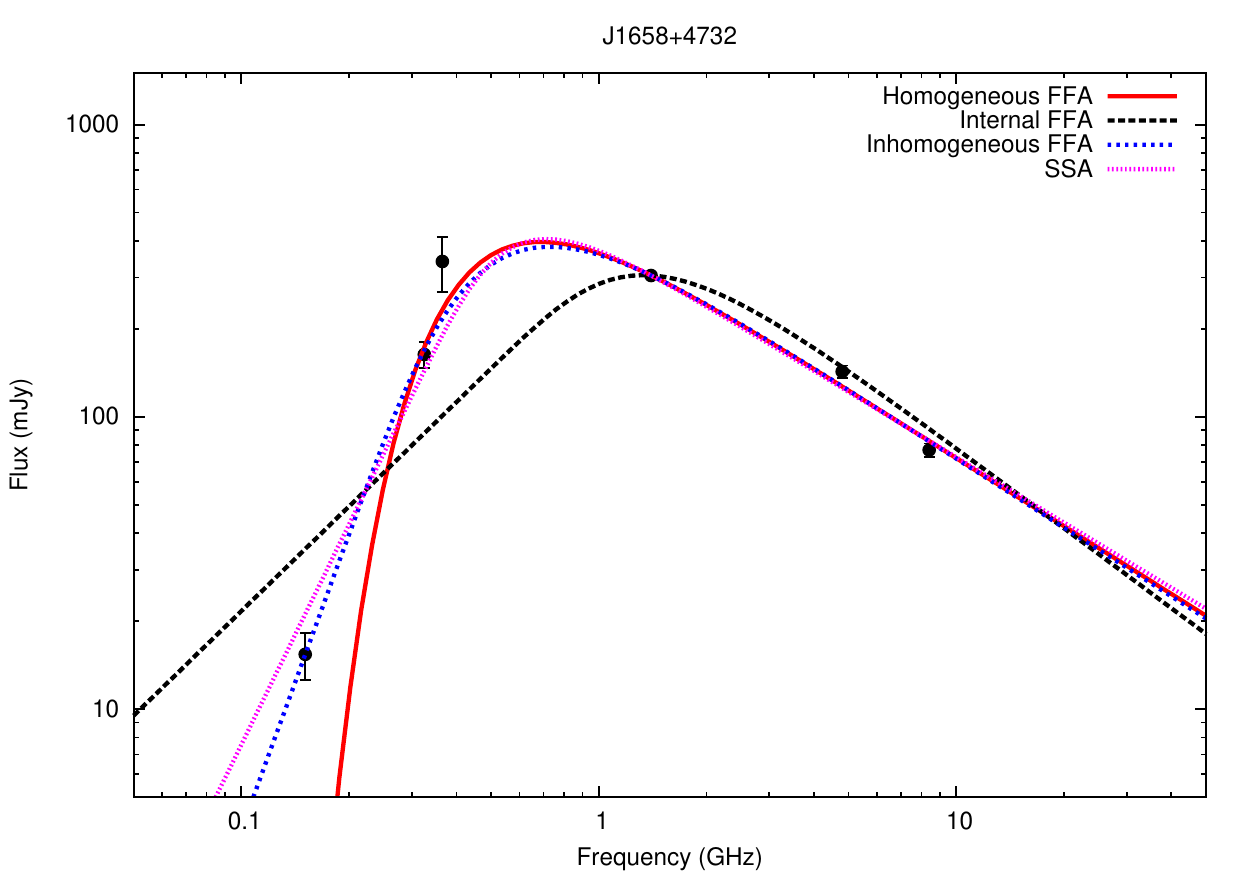}
\caption{Different absorption models fitted to the spectral energy distributions of the bonafide EISERS J1326$+$5712 and J1658$+$4732. In both cases the inhomogeneous FFA model provides the best fit (see Section5).} 
\label{fig:fitting}  
\end{figure*}

From Table~\ref{table:fitting}, it can be seen that the inhomogeneous FFA model by \citet{Bicknell1997} is preferred over other models, for 8 out of the 15 sources. For 2 of the remaining 7 sources, the SSA model provides the best fit. None of the models gave a satisfactory fit for the remaining 5 sources, either due to lack of sufficient data or due to the relatively flat spectral shape. Here, it should be noted that for 3 of the best EISERS candidates (Section \ref{section6}), namely, J0858$+$7501, J1326$+$5712 and J1658$+$4732, the best fitting model is SSA in one case and inhomogeneous FFA in the other two cases of confirmed EISERS. The various absorption models fitted to the spectra of the two confirmed EISERS are shown in Figure~\ref{fig:fitting}.

However, we caution that the above inference is only suggestive, in view of the scarcity of flux density measurements in the inverted part of the radio spectrum. To distinctly identify the best representative absorption model with a high statistical confidence, more data points will be needed below the turnover frequency.  

\subsection{Measuring Physical Parameters}
For the confirmed EISERS where in-homogeneous FFA model by \citet{Bicknell1997} provides the best fit, namely, J0847$+$5723, J1326$+$5712 and J1658$+$4732, we estimate the typical densities of the absorbing medium required to obtain an optical depth of unity at the turnover frequency. We use analytical models given in \citet{Bicknell2018} to calculate the specific intensity of synchrotron emitting plasma embedded in a clumpy free-free absorbing medium where the density of the ionized medium follows a log-normal distribution. More details can be found in Section 5.5 in \citet{Zovaro2019}. Here, we assume the maximum depth of absorbing screen to be the same as the extent of the source, which is obtained from the VLBI images taken from Astrogeo catalog. We obtain the peak frequency from spectral fitting of \citet{Bicknell1997} model for FFA. The velocity dispersion measurement of the absorbing medium for these sources could not be confirmed due to lack of optical observations. Following \citet{Schreiber2009} and \citet{Mukherjee2016} we assume a typical Baryonic velocity dispersion of 250 km s$^{-1}$. The values of mean densities thus obtained for J0847$+$5723, J1326$+$5712 and J1658$+$4732 are 36.2$\pm$11.6, 43.2$\pm$5.3 and 17.11$\pm$2.2 cm$^{-3}$ respectively. These mean densities of the absorbing thermal gas match well with the values predicted in the theoretical models for jet propagation in ionized medium by \citet{Begelman1996} and \citet{Bicknell1997}. \\

For sources in which SSA is the best fitting spectral model, the theoretical relation between magnetic field, turnover frequency and angular size of a source \citep[e.g.][]{Kellermann1969} can provide an estimate of magnetic field strength. For the sources J0858$+$7501 and J1700$+$3830, the magnetic field thus calculated is 4.1 G and 4.2 G respectively. This is an order of magnitude larger than the typical magnetic field strength of 5 to 100 mG reported for GPS sources \citep{Odea1998,Orienti2008}. However, for the similarly extreme low frequency spectral index source, PKS B0008-421, \citet{Callingham2015} have again found a high value, B $\sim$ 4.1 G.

\begin{table}
\small
\caption{Best-fit models for the present sample of sources showing spectral turnover} 
\label{table:fitting}
\begin{tabular}{cccc}\\
\hline
\multicolumn{1}{c}{Source} & \multicolumn{1}{c}{$\alpha^{a}$} 
& \multicolumn{1}{c}{Best fit model}& \multicolumn{1}{c}{$\nu_{peak}^{b}$ (GHz)}\\
\hline
J0045$+$8810 &1.41${\pm}$0.25 &Insufficient data& -\\
J0304$+$7727 &2.13${\pm}$0.20 &Inhomogeneous FFA&1.08\\
J0754$+$5324 &1.83${\pm}$0.22 &Inhomogeneous FFA&1.47\\
\textbf{J0847$+$5723} &2.36${\pm}$0.25 &Inhomogeneous FFA&0.61\\
\textbf{J0858$+$7501} &2.40${\pm}$0.34 &SSA&0.94\\
\textbf{J1326$+$5712} &2.91${\pm}$0.20 &Inhomogeneous FFA&0.73\\ 
J1430$+$3649 &0.58${\pm}$0.26 &Flat spectrum&-\\
J1536$+$8154 &2.24${\pm}$0.25 &Insufficient data&-\\
\textbf{J1549$+$5038} &2.42${\pm}$0.22 &Inhomogeneous FFA&0.42\\
\textbf{J1658$+$4732} &3.05${\pm}$0.27 &Inhomogeneous FFA&0.57\\
J1700$+$3830 &1.64${\pm}$0.21 &SSA&0.45\\
J1722$+$7046 &1.55${\pm}$0.22 &Inhomogeneous FFA&0.75\\
J1723$+$7653 &0.43${\pm}$0.18 &Flat-spectrum&-\\
J1846$+$4239 &1.48${\pm}$0.21 &Inhomogeneous FFA&1.08\\
J2317$+$4738 &2.05${\pm}$0.21 &Insufficient data&-\\
\hline
\end{tabular}

{$^{a}$ Spectral Index based on the 150 MHz and 325 MHz flux densities measured from the present uGMRT maps.}\\ 
{$^{b}$ Based on the best fit model.}\\
{The bonafide and tentative EISERS are marked in bold.}\\
\end{table}

\section{Discussion}\label{section6}
From Table~\ref{table:summary} it is seen that out of the 15 EISERS candidates observed in this study, J1326$+$5712 and J1658$+$4732 are confirmed as bona-fide EISERS with sharply inverted radio spectra between 150 MHz and 325 MHz. Based on our quasi-simultaneous uGMRT observations at these two frequencies, we find their spectral indices to be $\alpha$ (150 - 325 MHz) = $+$2.91${\pm}$0.20 and $+$3.05${\pm}$0.27, respectively (Table~\ref{table:spec-prop}), which are significantly above the critical limit of $\alpha_c$ = $+$ 2.5 for the standard case of synchrotron self-absorption (SSA, Section \ref{section1}). Both sources have a GPS type spectrum peaked near 1 GHz (Figure~\ref{fig:spec_all}). Unfortunately, their optical counterparts are presently unknown. 

Further, as seen from Table~\ref{table:spec-prop}, spectral indices of another 3 sources are consistent with $\alpha_c$ = $+$ 2.5, to within  1$\sigma$ uncertainty. We designate them as `possible' EISERS. These sources are: J0847$+$5723 ($\alpha$ = $+$2.36${\pm}$0.25), J0858$+$7501 ($\alpha$ = $+$2.40${\pm}$0.34) and J1549$+$5038 ($\alpha$ = $+$2.42${\pm}$0.22). It is reiterated that the spectral indices of the two confirmed EISERS would remain above the SSA limit of $\alpha_c$ = +2.5 even if their present flux densities measured at 150 MHz with uGMRT are aligned with the flux scale of the TGSS-ADR1 (Section 3.2). The radio power at 5 GHz are presentented in Table~\ref{table:summary}. Since redshifts are not available for most of the EISERS candidates observed here,  we have assumed a redshift z=1 for calculating the radio powers. All the 15 observed EISERS candidates are above the FR1/FR2 break power of $\sim$ 10$^{25}$W Hz$^{-1}$ and the 5 GHz radio powers of the confirmed/tentative EISERS are similar to the GPS/CSS type sources \citep{Odea1998}.

From Figure~\ref{fig:spec_all}, at least 9 of the 15 sources are seen to have a GPS type radio spectrum marked by a single peak in the frequency range 0.3 -- 1.2 GHz. Consequently, even their metre-wavelength emission should originate predominantly in milli-arcsecond components. Although GPS sources (when identified with galaxies) are usually only mildly variable at centimetre wavelengths where they become transparent \citep[e.g.][]{Stanghellini2001}, they are compact enough to be prone to showing flux-density modulations at metre wavelengths \citep{Bell2019}. This is most likely due to refractive interstellar scintillations (RISS) from the turbulent ionised medium within our galaxy \citep{Rickett1984, Jauncey2016, Shapirovskaya1978}. Thus, it is important that their radio spectral indices are based on multi-frequency observations made at least quasi-simultaneously. The present uGMRT observations of our 15 candidate EISERS at 150 MHz and 325 MHz were separated by less than a day, the only exception being the source J1658+4732 ($\alpha$ (150$-$325 MHz) = $+$3.05$\pm$0.27), for which the time interval was 16 days, due to scheduling problems (Table~\ref{table:obs-log}). However, for all practical purposes, even its observations can be regarded quasi-simultaneous, given that the expected time scale for RISS is year-like at 325 MHz and decade-like at 150 MHz \citep{Bell2019, 
Lazio2004}. For 3 sources in the sample, the radio spectra are found to clearly differ from the GPS type. These are J1430+3649, J1549+5038 and possibly J1723+7653. Their radio spectra exhibit evidence for a high opacity at least up to $\sim$ 10 GHz (Figure~\ref{fig:spec_all}), but do not resemble \lq High-Frequency Peakers (HFPs)\rq \citep[e.g.][]{Dallacasa2000,Orienti2007}.

Based on a recent study using the GLEAM survey \citep{Hurley-Walker2017}, \citet{Callingham2017} have reported a few radio sources 
having extremely inverted radio spectra. Two of them, J213024$-$434819 and J213024$-$434819 are promising candidates for radio 
sources in which the critical spectral slope ($\alpha_c$ $=$ $+$2.5) attainable in the SSA limit for a standard source of
incoherent synchrotron radiation is breached (see, however, the cautionary remarks in \citet{Mhaskey2019} about both these sources). 
On statistical grounds, their most favoured interpretation, based on the spectral modeling, is that these extreme radio sources 
(i.e., EISERS) represent cases whose synchrotron radio lobes are enveloped by a nearly homogeneous
screen of thermal plasma which is responsible for the steep spectral turnover via FFA effects \citep{Callingham2017}.
On the other hand, if the FFA interpretation remains unsubstantiated in cases of some EISERS, their ultra-steep spectral turnover 
would be indicative of a non-standard energy distribution of the relativistic particles, e.g., a delta-function or a Maxwellian 
\citep{Rees1967, Schlickeiser1985, Stawarz2008, Asano2014}, or, alternatively, of a large excess of electrons at lower energies, over the power-law extrapolated from higher 
energies \citep{deKool1989}. This underscores the importance of obtaining observational clues to verify the presence (or, the lack) 
of a sufficiently opaque sheath of thermal plasma enveloping most of the radio-emitting volume in EISERS.     

A preliminary spectral modelling performed on the EISERS candidates (Section \ref{section5.3}) underlines the relevance of the inhomogeneous FFA model \citep{Bicknell1997} in the absorption processes operating in these rare sources. A more promising approach towards modeling such sharp spectral turnovers may well become applicable in the near future \citep{Bicknell2018}.  

\section{Conclusions}\label{section7}
In this paper, we have reported on our ongoing search for extragalactic sources whose radio continuum spectrum turns over 
sharply due to absorption effects, so that the logarithmic slope of the inverted radio spectrum, $\alpha$, exceeds $+$2.5. 
Such a steep inversion of radio spectrum would be inconsistent with the standard explanation which invokes synchrotron 
self-absorption occurring in a source of incoherent synchrotron radiation emitted by relativistic electrons whose 
energy distribution follows a power-law (Section \ref{section1}). 

In this paper we have built upon the work published in our earlier two papers (Papers I \& II, see Section \ref{section1}) wherein one, but 
quite plausibly both EISERS candidates originally identified in Paper I as the most promising candidates known at that time, were 
confirmed in Paper II. 
In the present study, we have used the upgraded GMRT to carry out near-simultaneous observations with high resolution and 
sensitivity at 150 and 325 MHz, of 15 EISERS candidates in the northern sky, which we had short-listed by combining the TGSS-ADR1 (150 MHz) and WENSS 
(325 MHz) surveys. The present observations confirm at least 2 of these 15 candidates, namely J1326$+$5712 and J1658$+$4732, as
 bona fide EISERS. We find both to have an ultra-sharply inverted radio spectrum, with slopes of +2.91$\pm$0.20 and +3.05$\pm$0.27 
between 150 MHz and 325 MHz, respectively. Also, both have GPS type spectra peaking near 1 GHz. VLBI image of J1326$+$5712 
exhibits a CSO type morphology, while J1658$+$4732 exhibits a core-jet structure.

The new EISERS reported here represent a significant addition to the previously reported EISERS, namely J0242-1649 (confirmed
EISERS, Paper II), J1209$-$2032(probable EISERS, Paper II) and one possible case, J213024$-$434819 \citep[][but, see paper II]
{Callingham2017}. Their extreme rarity puts stringent constraints on the physical conditions in the peaked spectrum radio sources and makes them very interesting targets for multi-band observations, including optical spectroscopy, as well as radio spectroscopy (for detecting HI absorption) and radio polarimetry (for measuring the Rotation Measure). Such investigations, together with the theoretical modeling of their radio spectral turnover would play a critical role in deciding whether the ultra-sharp spectral turnover at low radio frequencies is caused primarily due to free-free absorption in thermal plasma, or whether it flags a non-standard particle acceleration mechanism operating in these rare extragalactic radio sources.


\clearpage
\newpage
\section*{Acknowledgements}
We thank the staff of the GMRT who have made these observations possible. GMRT is run by the National Centre for Radio Astrophysics 
of the Tata Institute of Fundamental Research. This research has used NASA's Astrophysics Data System and NASA/IPAC Extragalactic Database (NED), Jet Propulsion Laboratory, California 
Institute of Technology under contract with National Aeronautics and Space Administration and VizieR catalogue access tool, CDS, 
Strasbourg, France. SP would like to thank DST INSPIRE Faculty Scheme (IF12/PH-44) for funding his research group. We would like to thank Dipanjan Mukherjee, the referee to this manuscript, for his constructive comments.
\clearpage

\newpage
\bibliographystyle{mnras}
\bibliography{references}

\begin{thebibliography}{}
\makeatletter
\relax
\def\mn@urlcharsother{\let\do\@makeother \do\$\do\&\do\#\do\^\do\_\do\%\do\~}
\def\mn@doi{\begingroup\mn@urlcharsother \@ifnextchar [ {\mn@doi@}
  {\mn@doi@[]}}
\def\mn@doi@[#1]#2{\def\@tempa{#1}\ifx\@tempa\@empty \href
  {http://dx.doi.org/#2} {doi:#2}\else \href {http://dx.doi.org/#2} {#1}\fi
  \endgroup}
\def\mn@eprint#1#2{\mn@eprint@#1:#2::\@nil}
\def\mn@eprint@arXiv#1{\href {http://arxiv.org/abs/#1} {{\tt arXiv:#1}}}
\def\mn@eprint@dblp#1{\href {http://dblp.uni-trier.de/rec/bibtex/#1.xml}
  {dblp:#1}}
\def\mn@eprint@#1:#2:#3:#4\@nil{\def\@tempa {#1}\def\@tempb {#2}\def\@tempc
  {#3}\ifx \@tempc \@empty \let \@tempc \@tempb \let \@tempb \@tempa \fi \ifx
  \@tempb \@empty \def\@tempb {arXiv}\fi \@ifundefined
  {mn@eprint@\@tempb}{\@tempb:\@tempc}{\expandafter \expandafter \csname
  mn@eprint@\@tempb\endcsname \expandafter{\@tempc}}}

\bibitem[\protect\citeauthoryear{{Alam} et~al.,}{{Alam}
  et~al.}{2015}]{Alam2015}
{Alam} S.,  et~al., 2015, \mn@doi [\apjs] {10.1088/0067-0049/219/1/12}, \href
  {http://adsabs.harvard.edu/abs/2015ApJS..219...12A} {219, 12}

\bibitem[\protect\citeauthoryear{{An} \& {Baan}}{{An} \& {Baan}}{2012}]{An2012}
{An} T.,  {Baan} W.~A.,  2012, \mn@doi [\apj] {10.1088/0004-637X/760/1/77},
  \href {http://adsabs.harvard.edu/abs/2012ApJ...760...77A} {760, 77}

\bibitem[\protect\citeauthoryear{{Asano}, {Takahara}, {Kusunose}, {Toma}  \&
  {Kakuwa}}{{Asano} et~al.}{2014}]{Asano2014}
{Asano} K.,  {Takahara} F.,  {Kusunose} M.,  {Toma} K.,   {Kakuwa} J.,  2014,
  \mn@doi [\apj] {10.1088/0004-637X/780/1/64}, \href
  {https://ui.adsabs.harvard.edu/abs/2014ApJ...780...64A} {780, 64}

\bibitem[\protect\citeauthoryear{{Baars}, {Genzel}, {Pauliny-Toth}  \&
  {Witzel}}{{Baars} et~al.}{1977}]{Baars1977}
{Baars} J.~W.~M.,  {Genzel} R.,  {Pauliny-Toth} I.~I.~K.,   {Witzel} A.,  1977,
  \aap, \href {http://adsabs.harvard.edu/abs/1977A%26A....61...99B} {61, 99}

\bibitem[\protect\citeauthoryear{{Baldi}, {Capetti}  \& {Giovannini}}{{Baldi}
  et~al.}{2015}]{Baldi2015}
{Baldi} R.~D.,  {Capetti} A.,   {Giovannini} G.,  2015, \mn@doi [\aap]
  {10.1051/0004-6361/201425426}, \href
  {http://adsabs.harvard.edu/abs/2015A%26A...576A..38B} {576, A38}

\bibitem[\protect\citeauthoryear{{Begelman}}{{Begelman}}{1996}]{Begelman1996}
{Begelman} M.~C.,  1996, {Baby Cygnus A's}.
p.~209

\bibitem[\protect\citeauthoryear{{Bell} et~al.,}{{Bell}
  et~al.}{2019}]{Bell2019}
{Bell} M.~E.,  et~al., 2019, \mn@doi [\mnras] {10.1093/mnras/sty2801}, \href
  {http://adsabs.harvard.edu/abs/2019MNRAS.482.2484B} {482, 2484}

\bibitem[\protect\citeauthoryear{{Best}, {Kauffmann}, {Heckman}  \&
  {Ivezi{\'c}}}{{Best} et~al.}{2005}]{Best2005}
{Best} P.~N.,  {Kauffmann} G.,  {Heckman} T.~M.,   {Ivezi{\'c}} {\v Z}.,  2005,
  \mn@doi [\mnras] {10.1111/j.1365-2966.2005.09283.x}, \href
  {http://adsabs.harvard.edu/abs/2005MNRAS.362....9B} {362, 9}

\bibitem[\protect\citeauthoryear{{Bicknell}, {Dopita}  \& {O'Dea}}{{Bicknell}
  et~al.}{1997}]{Bicknell1997}
{Bicknell} G.~V.,  {Dopita} M.~A.,   {O'Dea} C.~P.~O.,  1997, \mn@doi [\apj]
  {10.1086/304400}, \href {http://adsabs.harvard.edu/abs/1997ApJ...485..112B}
  {485, 112}

\bibitem[\protect\citeauthoryear{{Bicknell}, {Mukherjee}, {Wagner},
  {Sutherland}  \& {Nesvadba}}{{Bicknell} et~al.}{2018}]{Bicknell2018}
{Bicknell} G.~V.,  {Mukherjee} D.,  {Wagner} A.~Y.,  {Sutherland} R.~S.,
  {Nesvadba} N.~P.~H.,  2018, \mn@doi [\mnras] {10.1093/mnras/sty070}, \href
  {http://adsabs.harvard.edu/abs/2018MNRAS.475.3493B} {475, 3493}

\bibitem[\protect\citeauthoryear{{Callingham} et~al.,}{{Callingham}
  et~al.}{2015}]{Callingham2015}
{Callingham} J.~R.,  et~al., 2015, \mn@doi [\apj]
  {10.1088/0004-637X/809/2/168}, \href
  {http://adsabs.harvard.edu/abs/2015ApJ...809..168C} {809, 168}

\bibitem[\protect\citeauthoryear{{Callingham} et~al.,}{{Callingham}
  et~al.}{2017}]{Callingham2017}
{Callingham} J.~R.,  et~al., 2017, \mn@doi [\apj]
  {10.3847/1538-4357/836/2/174}, \href
  {http://adsabs.harvard.edu/abs/2017ApJ...836..174C} {836, 174}

\bibitem[\protect\citeauthoryear{{Chandra}, {Ray}  \& {Bhatnagar}}{{Chandra}
  et~al.}{2004}]{Chandra2004}
{Chandra} P.,  {Ray} A.,   {Bhatnagar} S.,  2004, \mn@doi [\apj]
  {10.1086/422675}, \href {http://adsabs.harvard.edu/abs/2004ApJ...612..974C}
  {612, 974}

\bibitem[\protect\citeauthoryear{{Cheung} et~al.,}{{Cheung}
  et~al.}{2013}]{Cheung2013}
{Cheung} E.,  et~al., 2013, \mn@doi [\apj] {10.1088/0004-637X/779/2/162}, \href
  {http://adsabs.harvard.edu/abs/2013ApJ...779..162C} {779, 162}

\bibitem[\protect\citeauthoryear{{Condon}, {Cotton}, {Greisen}, {Yin},
  {Perley}, {Taylor}  \& {Broderick}}{{Condon} et~al.}{1998}]{Condon1998}
{Condon} J.~J.,  {Cotton} W.~D.,  {Greisen} E.~W.,  {Yin} Q.~F.,  {Perley}
  R.~A.,  {Taylor} G.~B.,   {Broderick} J.~J.,  1998, \mn@doi [\aj]
  {10.1086/300337}, \href {http://adsabs.harvard.edu/abs/1998AJ....115.1693C}
  {115, 1693}

\bibitem[\protect\citeauthoryear{{Conway}, {Myers}, {Pearson}, {Readhead},
  {Unwin}  \& {Xu}}{{Conway} et~al.}{1994}]{Conway1994}
{Conway} J.~E.,  {Myers} S.~T.,  {Pearson} T.~J.,  {Readhead} A.~C.~S.,
  {Unwin} S.~C.,   {Xu} W.,  1994, \mn@doi [\apj] {10.1086/174006}, \href
  {http://adsabs.harvard.edu/abs/1994ApJ...425..568C} {425, 568}

\bibitem[\protect\citeauthoryear{{Coziol}, {Andernach}, {Torres-Papaqui},
  {Ortega-Minakata}  \& {Moreno del Rio}}{{Coziol} et~al.}{2017}]{Coziol2017}
{Coziol} R.,  {Andernach} H.,  {Torres-Papaqui} J.~P.,  {Ortega-Minakata}
  R.~A.,   {Moreno del Rio} F.,  2017, \mn@doi [\mnras]
  {10.1093/mnras/stw3164}, \href
  {http://adsabs.harvard.edu/abs/2017MNRAS.466..921C} {466, 921}

\bibitem[\protect\citeauthoryear{{Dabhade}, {Gaikwad}, {Bagchi},
  {Pandey-Pommier}, {Sankhyayan}  \& {Raychaudhury}}{{Dabhade}
  et~al.}{2017}]{Dabhade2017}
{Dabhade} P.,  {Gaikwad} M.,  {Bagchi} J.,  {Pandey-Pommier} M.,  {Sankhyayan}
  S.,   {Raychaudhury} S.,  2017, \mn@doi [\mnras] {10.1093/mnras/stx860},
  \href {http://adsabs.harvard.edu/abs/2017MNRAS.469.2886D} {469, 2886}

\bibitem[\protect\citeauthoryear{{Dallacasa}, {Stanghellini}, {Centonza}  \&
  {Fanti}}{{Dallacasa} et~al.}{2000}]{Dallacasa2000}
{Dallacasa} D.,  {Stanghellini} C.,  {Centonza} M.,   {Fanti} R.,  2000, \aap,
  \href {http://adsabs.harvard.edu/abs/2000A%26A...363..887D} {363, 887}

\bibitem[\protect\citeauthoryear{{De Breuck}, {Tang}, {de Bruyn},
  {R{\"o}ttgering}  \& {van Breugel}}{{De Breuck} et~al.}{2002}]{DeBreuck2002}
{De Breuck} C.,  {Tang} Y.,  {de Bruyn} A.~G.,  {R{\"o}ttgering} H.,   {van
  Breugel} W.,  2002, \mn@doi [\aap] {10.1051/0004-6361:20021115}, \href
  {http://adsabs.harvard.edu/abs/2002A%26A...394...59D} {394, 59}

\bibitem[\protect\citeauthoryear{{Douglas}, {Bash}, {Bozyan}, {Torrence}  \&
  {Wolfe}}{{Douglas} et~al.}{1996}]{Douglas1996}
{Douglas} J.~N.,  {Bash} F.~N.,  {Bozyan} F.~A.,  {Torrence} G.~W.,   {Wolfe}
  C.,  1996, \mn@doi [\aj] {10.1086/117932}, \href
  {http://adsabs.harvard.edu/abs/1996AJ....111.1945D} {111, 1945}

\bibitem[\protect\citeauthoryear{{Fey} \& {Charlot}}{{Fey} \&
  {Charlot}}{2000}]{Fey2000}
{Fey} A.~L.,  {Charlot} P.,  2000, \mn@doi [\apjs] {10.1086/313382}, \href
  {http://adsabs.harvard.edu/abs/2000ApJS..128...17F} {128, 17}

\bibitem[\protect\citeauthoryear{{F{\"o}rster Schreiber} et~al.,}{{F{\"o}rster
  Schreiber} et~al.}{2009}]{Schreiber2009}
{F{\"o}rster Schreiber} N.~M.,  et~al., 2009, \mn@doi [\apj]
  {10.1088/0004-637X/706/2/1364}, \href
  {https://ui.adsabs.harvard.edu/abs/2009ApJ...706.1364F} {706, 1364}

\bibitem[\protect\citeauthoryear{{Gopal-Krishna} \&
  {Spoelstra}}{{Gopal-Krishna} \& {Spoelstra}}{1993}]{GopalKrishna1993}
{Gopal-Krishna} {Spoelstra} T.~A.~T.,  1993, \aap, \href
  {http://adsabs.harvard.edu/abs/1993A%26A...271..101G} {271, 101}

\bibitem[\protect\citeauthoryear{{Gopal-Krishna} \& {Wiita}}{{Gopal-Krishna} \&
  {Wiita}}{2000}]{Gopal-Krishna2000}
{Gopal-Krishna} {Wiita} P.~J.,  2000, \aap, \href
  {http://adsabs.harvard.edu/abs/2000A%26A...363..507G} {363, 507}

\bibitem[\protect\citeauthoryear{{Gopal-Krishna}, {Patnaik}  \&
  {Steppe}}{{Gopal-Krishna} et~al.}{1983}]{GopalKrishna1983}
{Gopal-Krishna} {Patnaik} A.~R.,   {Steppe} H.,  1983, \aap, \href
  {http://adsabs.harvard.edu/abs/1983A%26A...123..107G} {123, 107}

\bibitem[\protect\citeauthoryear{{Gopal-Krishna}, {Biermann}, {Gergely}  \&
  {Wiita}}{{Gopal-Krishna} et~al.}{2012}]{GK2012}
{Gopal-Krishna} {Biermann} P.~L.,  {Gergely} L.~{\'A}.,   {Wiita} P.~J.,  2012,
  \mn@doi [Research in Astronomy and Astrophysics]
  {10.1088/1674-4527/12/2/002}, \href
  {http://adsabs.harvard.edu/abs/2012RAA....12..127G} {12, 127}

\bibitem[\protect\citeauthoryear{{Gopal-Krishna}, {Sirothia}, {Mhaskey},
  {Ranadive}, {Wiita}, {Goyal}, {Kantharia}  \&
  {Ishwara-Chandra}}{{Gopal-Krishna} et~al.}{2014}]{GopalKrishna2014}
{Gopal-Krishna} {Sirothia} S.~K.,  {Mhaskey} M.,  {Ranadive} P.,  {Wiita}
  P.~J.,  {Goyal} A.,  {Kantharia} N.~G.,   {Ishwara-Chandra} C.~H.,  2014,
  \mn@doi [\mnras] {10.1093/mnras/stu1364}, \href
  {http://adsabs.harvard.edu/abs/2014MNRAS.443.2824K} {443, 2824}

\bibitem[\protect\citeauthoryear{{Gregory} \& {Condon}}{{Gregory} \&
  {Condon}}{1991}]{Gregory1991}
{Gregory} P.~C.,  {Condon} J.~J.,  1991, \mn@doi [\apjs] {10.1086/191559},
  \href {http://adsabs.harvard.edu/abs/1991ApJS...75.1011G} {75, 1011}

\bibitem[\protect\citeauthoryear{{Gugliucci}, {Taylor}, {Peck}  \&
  {Giroletti}}{{Gugliucci} et~al.}{2005}]{Gugliucci2005}
{Gugliucci} N.~E.,  {Taylor} G.~B.,  {Peck} A.~B.,   {Giroletti} M.,  2005,
  \mn@doi [\apj] {10.1086/427934}, \href
  {http://adsabs.harvard.edu/abs/2005ApJ...622..136G} {622, 136}

\bibitem[\protect\citeauthoryear{Gupta et~al.,}{Gupta et~al.}{2017}]{Gupta2017}
Gupta Y.,  et~al., 2017, Current Science, 113, 707

\bibitem[\protect\citeauthoryear{{Hales}, {Mayer}, {Warner}  \&
  {Baldwin}}{{Hales} et~al.}{1991}]{Hales1991}
{Hales} S.~E.~G.,  {Mayer} C.~J.,  {Warner} P.~J.,   {Baldwin} J.~E.,  1991,
  \mn@doi [\mnras] {10.1093/mnras/251.1.46}, \href
  {https://ui.adsabs.harvard.edu/abs/1991MNRAS.251...46H} {251, 46}

\bibitem[\protect\citeauthoryear{{Hardcastle} et~al.,}{{Hardcastle}
  et~al.}{2016}]{Hardcastle2016}
{Hardcastle} M.~J.,  et~al., 2016, \mn@doi [\mnras] {10.1093/mnras/stw1763},
  \href {http://adsabs.harvard.edu/abs/2016MNRAS.462.1910H} {462, 1910}

\bibitem[\protect\citeauthoryear{{Helfand}, {White}  \& {Becker}}{{Helfand}
  et~al.}{2015}]{Helfand2015}
{Helfand} D.~J.,  {White} R.~L.,   {Becker} R.~H.,  2015, \mn@doi [\apj]
  {10.1088/0004-637X/801/1/26}, \href
  {http://adsabs.harvard.edu/abs/2015ApJ...801...26H} {801, 26}

\bibitem[\protect\citeauthoryear{{Helmboldt} et~al.,}{{Helmboldt}
  et~al.}{2007}]{Helmboldt2007}
{Helmboldt} J.~F.,  et~al., 2007, \mn@doi [\apj] {10.1086/511005}, \href
  {http://adsabs.harvard.edu/abs/2007ApJ...658..203H} {658, 203}

\bibitem[\protect\citeauthoryear{{Hurley-Walker} et~al.,}{{Hurley-Walker}
  et~al.}{2017}]{Hurley-Walker2017}
{Hurley-Walker} N.,  et~al., 2017, \mn@doi [\mnras] {10.1093/mnras/stw2337},
  \href {http://adsabs.harvard.edu/abs/2017MNRAS.464.1146H} {464, 1146}

\bibitem[\protect\citeauthoryear{{Intema}}{{Intema}}{2014}]{Intema2014}
{Intema} H.~T.,  2014, in Astronomical Society of India Conference Series.
  (\mn@eprint {arXiv} {1402.4889})

\bibitem[\protect\citeauthoryear{{Intema}, {Jagannathan}, {Mooley}  \&
  {Frail}}{{Intema} et~al.}{2017}]{Intema2017}
{Intema} H.~T.,  {Jagannathan} P.,  {Mooley} K.~P.,   {Frail} D.~A.,  2017,
  \mn@doi [\aap] {10.1051/0004-6361/201628536}, \href
  {http://adsabs.harvard.edu/abs/2017A%26A...598A..78I} {598, A78}

\bibitem[\protect\citeauthoryear{{Jauncey} et~al.,}{{Jauncey}
  et~al.}{2016}]{Jauncey2016}
{Jauncey} D.,  et~al., 2016, \mn@doi [Galaxies] {10.3390/galaxies4040062},
  \href {http://adsabs.harvard.edu/abs/2016Galax...4...62J} {4, 62}

\bibitem[\protect\citeauthoryear{{Jones} et~al.,}{{Jones}
  et~al.}{1996}]{Jones1996}
{Jones} D.~L.,  et~al., 1996, \mn@doi [\apjl] {10.1086/310183}, \href
  {http://adsabs.harvard.edu/abs/1996ApJ...466L..63J} {466, L63}

\bibitem[\protect\citeauthoryear{{Kameno}, {Horiuchi}, {Shen}, {Inoue},
  {Kobayashi}, {Hirabayashi}  \& {Murata}}{{Kameno} et~al.}{2000}]{Kameno2000}
{Kameno} S.,  {Horiuchi} S.,  {Shen} Z.-Q.,  {Inoue} M.,  {Kobayashi} H.,
  {Hirabayashi} H.,   {Murata} Y.,  2000, \mn@doi [\pasj]
  {10.1093/pasj/52.1.209}, \href
  {http://adsabs.harvard.edu/abs/2000PASJ...52..209K} {52, 209}

\bibitem[\protect\citeauthoryear{{Kapi{\'n}ska} et~al.,}{{Kapi{\'n}ska}
  et~al.}{2017}]{Kapinska2017}
{Kapi{\'n}ska} A.~D.,  et~al., 2017, \mn@doi [\aj] {10.3847/1538-3881/aa90b7},
  \href {http://adsabs.harvard.edu/abs/2017AJ....154..253K} {154, 253}

\bibitem[\protect\citeauthoryear{{Kellermann}}{{Kellermann}}{1966}]{Kellermann1966}
{Kellermann} K.~I.,  1966, \mn@doi [Australian Journal of Physics]
  {10.1071/PH660195}, \href {http://adsabs.harvard.edu/abs/1966AuJPh..19..195K}
  {19, 195}

\bibitem[\protect\citeauthoryear{{Kellermann} \& {Pauliny-Toth}}{{Kellermann}
  \& {Pauliny-Toth}}{1969}]{Kellermann1969}
{Kellermann} K.~I.,  {Pauliny-Toth} I.~I.~K.,  1969, \mn@doi [\apjl]
  {10.1086/180305}, \href {http://adsabs.harvard.edu/abs/1969ApJ...155L..71K}
  {155, L71}

\bibitem[\protect\citeauthoryear{{Krichbaum}, {Alef}, {Witzel}, {Zensus},
  {Booth}, {Greve}  \& {Rogers}}{{Krichbaum} et~al.}{1998}]{Krichbaum1998}
{Krichbaum} T.~P.,  {Alef} W.,  {Witzel} A.,  {Zensus} J.~A.,  {Booth} R.~S.,
  {Greve} A.,   {Rogers} A.~E.~E.,  1998, \aap, \href
  {http://adsabs.harvard.edu/abs/1998A%26A...329..873K} {329, 873}

\bibitem[\protect\citeauthoryear{{Kuncic}, {Bicknell}  \& {Dopita}}{{Kuncic}
  et~al.}{1998}]{Kuncic1998}
{Kuncic} Z.,  {Bicknell} G.~V.,   {Dopita} M.~A.,  1998, \mn@doi [\apjl]
  {10.1086/311202}, \href {http://adsabs.harvard.edu/abs/1998ApJ...495L..35K}
  {495, L35}

\bibitem[\protect\citeauthoryear{{Ku{\'z}micz}, {Jamrozy},
  {Kozie{\l}-Wierzbowska}  \& {We{\.z}gowiec}}{{Ku{\'z}micz}
  et~al.}{2017}]{Kuzmicz2017}
{Ku{\'z}micz} A.,  {Jamrozy} M.,  {Kozie{\l}-Wierzbowska} D.,   {We{\.z}gowiec}
  M.,  2017, \mn@doi [\mnras] {10.1093/mnras/stx1830}, \href
  {http://adsabs.harvard.edu/abs/2017MNRAS.471.3806K} {471, 3806}

\bibitem[\protect\citeauthoryear{{Lane}, {Cotton}, {van Velzen}, {Clarke},
  {Kassim}, {Helmboldt}, {Lazio}  \& {Cohen}}{{Lane} et~al.}{2014}]{Lane2014}
{Lane} W.~M.,  {Cotton} W.~D.,  {van Velzen} S.,  {Clarke} T.~E.,  {Kassim}
  N.~E.,  {Helmboldt} J.~F.,  {Lazio} T.~J.~W.,   {Cohen} A.~S.,  2014, \mn@doi
  [\mnras] {10.1093/mnras/stu256}, \href
  {http://adsabs.harvard.edu/abs/2014MNRAS.440..327L} {440, 327}

\bibitem[\protect\citeauthoryear{{Lazio}, {Cordes}, {de Bruyn}  \&
  {Macquart}}{{Lazio} et~al.}{2004}]{Lazio2004}
{Lazio} T.~J.~W.,  {Cordes} J.~M.,  {de Bruyn} A.~G.,   {Macquart} J.-P.,
  2004, \mn@doi [\nar] {10.1016/j.newar.2004.09.039}, \href
  {http://adsabs.harvard.edu/abs/2004NewAR..48.1439L} {48, 1439}

\bibitem[\protect\citeauthoryear{{Levinson}, {Laor}  \& {Vermeulen}}{{Levinson}
  et~al.}{1995}]{Levinson1995}
{Levinson} A.,  {Laor} A.,   {Vermeulen} R.~C.,  1995, \mn@doi [\apj]
  {10.1086/175988}, \href {http://adsabs.harvard.edu/abs/1995ApJ...448..589L}
  {448, 589}

\bibitem[\protect\citeauthoryear{{Lowe}, {Gawro{\'n}ski}, {Wilkinson}, {Kus},
  {Browne}, {Pazderski}, {Feiler}  \& {Kettle}}{{Lowe} et~al.}{2007}]{Lowe2007}
{Lowe} S.~R.,  {Gawro{\'n}ski} M.~P.,  {Wilkinson} P.~N.,  {Kus} A.~J.,
  {Browne} I.~W.~A.,  {Pazderski} E.,  {Feiler} R.,   {Kettle} D.,  2007,
  \mn@doi [\aap] {10.1051/0004-6361:20078034}, \href
  {http://adsabs.harvard.edu/abs/2007A%26A...474.1093L} {474, 1093}

\bibitem[\protect\citeauthoryear{{Matveenko}, {Pauliny-Toth}  \&
  {Sherwood}}{{Matveenko} et~al.}{1990}]{Matveenko1990}
{Matveenko} L.~I.,  {Pauliny-Toth} I.~I.~K.,   {Sherwood} W.,  1990, Soviet
  Astronomy Letters, \href {http://adsabs.harvard.edu/abs/1990SvAL...16..247M}
  {16, 247}

\bibitem[\protect\citeauthoryear{{Mhaskey}, {Gopal-Krishna}, {Paul}, {Salunkhe}
   \& {Sirothia}}{{Mhaskey} et~al.}{2019}]{Mhaskey2019}
{Mhaskey} M.,  {Gopal-Krishna} Pratik D.,  {Paul} S.,  {Salunkhe} S.,
  {Sirothia} S.~K.,  2019, \mn@doi [\mnras] {10.1093/mnras/stz335}, \href
  {http://adsabs.harvard.edu/abs/2019MNRAS.tmp..341M} {}

\bibitem[\protect\citeauthoryear{{Mukherjee}, {Bicknell}, {Sutherland}  \&
  {Wagner}}{{Mukherjee} et~al.}{2016}]{Mukherjee2016}
{Mukherjee} D.,  {Bicknell} G.~V.,  {Sutherland} R.,   {Wagner} A.,  2016,
  \mn@doi [\mnras] {10.1093/mnras/stw1368}, \href
  {https://ui.adsabs.harvard.edu/abs/2016MNRAS.461..967M} {461, 967}

\bibitem[\protect\citeauthoryear{{Myers} et~al.,}{{Myers}
  et~al.}{2003}]{Myers2003}
{Myers} S.~T.,  et~al., 2003, VizieR Online Data Catalog, \href
  {http://adsabs.harvard.edu/abs/2003yCat.8072....0M} {8072}

\bibitem[\protect\citeauthoryear{{O'Dea}}{{O'Dea}}{1998}]{Odea1998}
{O'Dea} C.~P.,  1998, \mn@doi [\pasp] {10.1086/316162}, \href
  {http://adsabs.harvard.edu/abs/1998PASP..110..493O} {110, 493}

\bibitem[\protect\citeauthoryear{{Orienti} \& {Dallacasa}}{{Orienti} \&
  {Dallacasa}}{2008}]{Orienti2008}
{Orienti} M.,  {Dallacasa} D.,  2008, \mn@doi [\aap]
  {10.1051/0004-6361:20078098}, \href
  {http://adsabs.harvard.edu/abs/2008A%26A...477..807O} {477, 807}

\bibitem[\protect\citeauthoryear{{Orienti}, {Dallacasa}  \&
  {Stanghellini}}{{Orienti} et~al.}{2007}]{Orienti2007}
{Orienti} M.,  {Dallacasa} D.,   {Stanghellini} C.,  2007, \mn@doi [\aap]
  {10.1051/0004-6361:20078105}, \href
  {https://ui.adsabs.harvard.edu/abs/2007A&A...475..813O} {475, 813}

\bibitem[\protect\citeauthoryear{{Pacholczyk}}{{Pacholczyk}}{1970}]{Pacholczyk1970}
{Pacholczyk} A.~G.,  1970, {Radio astrophysics. Nonthermal processes in
  galactic and extragalactic sources}

\bibitem[\protect\citeauthoryear{{P{\^a}ris} et~al.,}{{P{\^a}ris}
  et~al.}{2017}]{Paris2017}
{P{\^a}ris} I.,  et~al., 2017, \mn@doi [\aap] {10.1051/0004-6361/201527999},
  \href {http://adsabs.harvard.edu/abs/2017A%26A...597A..79P} {597, A79}

\bibitem[\protect\citeauthoryear{{Pasetto}, {Kraus}, {Mack}, {Bruni}  \&
  {Carrasco-Gonz{\'a}lez}}{{Pasetto} et~al.}{2016}]{Pasetto2016}
{Pasetto} A.,  {Kraus} A.,  {Mack} K.-H.,  {Bruni} G.,
  {Carrasco-Gonz{\'a}lez} C.,  2016, \mn@doi [\aap]
  {10.1051/0004-6361/201526963}, \href
  {http://adsabs.harvard.edu/abs/2016A%26A...586A.117P} {586, A117}

\bibitem[\protect\citeauthoryear{{Peck} \& {Taylor}}{{Peck} \&
  {Taylor}}{2000}]{Peck2000}
{Peck} A.~B.,  {Taylor} G.~B.,  2000, \mn@doi [\apj] {10.1086/308746}, \href
  {http://adsabs.harvard.edu/abs/2000ApJ...534...90P} {534, 90}

\bibitem[\protect\citeauthoryear{{Phillips} \& {Mutel}}{{Phillips} \&
  {Mutel}}{1982}]{Phillips1982}
{Phillips} R.~B.,  {Mutel} R.~L.,  1982, \aap, \href
  {http://adsabs.harvard.edu/abs/1982A%26A...106...21P} {106, 21}

\bibitem[\protect\citeauthoryear{{Rees}}{{Rees}}{1967}]{Rees1967}
{Rees} M.~J.,  1967, \mn@doi [\mnras] {10.1093/mnras/136.3.279}, \href
  {http://adsabs.harvard.edu/abs/1967MNRAS.136..279R} {136, 279}

\bibitem[\protect\citeauthoryear{{Rengelink}, {Tang}, {de Bruyn}, {Miley},
  {Bremer}, {Roettgering}  \& {Bremer}}{{Rengelink}
  et~al.}{1997}]{Rengelink1997}
{Rengelink} R.~B.,  {Tang} Y.,  {de Bruyn} A.~G.,  {Miley} G.~K.,  {Bremer}
  M.~N.,  {Roettgering} H.~J.~A.,   {Bremer} M.~A.~R.,  1997, \mn@doi [\aaps]
  {10.1051/aas:1997358}, \href
  {http://adsabs.harvard.edu/abs/1997A%26AS..124..259R} {124, 259}

\bibitem[\protect\citeauthoryear{{Richards} et~al.,}{{Richards}
  et~al.}{2011}]{Richards2011}
{Richards} J.~L.,  et~al., 2011, \mn@doi [\apjs] {10.1088/0067-0049/194/2/29},
  \href {http://adsabs.harvard.edu/abs/2011ApJS..194...29R} {194, 29}

\bibitem[\protect\citeauthoryear{{Richards}, {Hovatta}, {Max-Moerbeck},
  {Pavlidou}, {Pearson}  \& {Readhead}}{{Richards} et~al.}{2014}]{Richards2014}
{Richards} J.~L.,  {Hovatta} T.,  {Max-Moerbeck} W.,  {Pavlidou} V.,  {Pearson}
  T.~J.,   {Readhead} A.~C.~S.,  2014, \mn@doi [\mnras]
  {10.1093/mnras/stt2412}, \href
  {http://adsabs.harvard.edu/abs/2014MNRAS.438.3058R} {438, 3058}

\bibitem[\protect\citeauthoryear{{Rickett}, {Coles}  \& {Bourgois}}{{Rickett}
  et~al.}{1984}]{Rickett1984}
{Rickett} B.~J.,  {Coles} W.~A.,   {Bourgois} G.,  1984, \aap, \href
  {http://adsabs.harvard.edu/abs/1984A%26A...134..390R} {134, 390}

\bibitem[\protect\citeauthoryear{{Roberts}, {Saripalli}, {Wang},
  {Sathyanarayana Rao}, {Subrahmanyan}, {KleinStern}, {Morii-Sciolla}  \&
  {Simpson}}{{Roberts} et~al.}{2018}]{Roberts2018}
{Roberts} D.~H.,  {Saripalli} L.,  {Wang} K.~X.,  {Sathyanarayana Rao} M.,
  {Subrahmanyan} R.,  {KleinStern} C.~C.,  {Morii-Sciolla} C.~Y.,   {Simpson}
  L.,  2018, \mn@doi [\apj] {10.3847/1538-4357/aa9c49}, \href
  {http://adsabs.harvard.edu/abs/2018ApJ...852...47R} {852, 47}

\bibitem[\protect\citeauthoryear{{Roger}, {Costain}  \& {Bridle}}{{Roger}
  et~al.}{1973}]{Roger1973}
{Roger} R.~S.,  {Costain} C.~H.,   {Bridle} A.~H.,  1973, \mn@doi [\aj]
  {10.1086/111506}, \href {http://adsabs.harvard.edu/abs/1973AJ.....78.1030R}
  {78, 1030}

\bibitem[\protect\citeauthoryear{{Rybicki} \& {Lightman}}{{Rybicki} \&
  {Lightman}}{1986}]{Rybicki1979}
{Rybicki} G.~B.,  {Lightman} A.~P.,  1986, {Radiative Processes in
  Astrophysics}

\bibitem[\protect\citeauthoryear{{Scheuer} \& {Williams}}{{Scheuer} \&
  {Williams}}{1968}]{Scheuer1968}
{Scheuer} P.~A.~G.,  {Williams} P.~J.~S.,  1968, \mn@doi [\araa]
  {10.1146/annurev.aa.06.090168.001541}, \href
  {http://adsabs.harvard.edu/abs/1968ARA%26A...6..321S} {6, 321}

\bibitem[\protect\citeauthoryear{{Schlickeiser}}{{Schlickeiser}}{1985}]{Schlickeiser1985}
{Schlickeiser} R.,  1985, {Particle acceleration in active galactic nuclei -
  coexistence of monoenergetic, power law and Maxwell-Boltzmann electron energy
  spectra.}.
pp 355--360

\bibitem[\protect\citeauthoryear{{Shapirovskaya}}{{Shapirovskaya}}{1978}]{Shapirovskaya1978}
{Shapirovskaya} N.~Y.,  1978, \sovast, \href
  {http://adsabs.harvard.edu/abs/1978SvA....22..544S} {22, 544}

\bibitem[\protect\citeauthoryear{{Shimwell} et~al.,}{{Shimwell}
  et~al.}{2017}]{Shimwell2017}
{Shimwell} T.~W.,  et~al., 2017, \mn@doi [\aap] {10.1051/0004-6361/201629313},
  \href {http://adsabs.harvard.edu/abs/2017A%26A...598A.104S} {598, A104}

\bibitem[\protect\citeauthoryear{{Shimwell} et~al.,}{{Shimwell}
  et~al.}{2019}]{Shimwell2019}
{Shimwell} T.~W.,  et~al., 2019, \mn@doi [\aap] {10.1051/0004-6361/201833559},
  \href {https://ui.adsabs.harvard.edu/abs/2019A&A...622A...1S} {622, A1}

\bibitem[\protect\citeauthoryear{{Slish}}{{Slish}}{1963}]{Slish1963}
{Slish} V.~I.,  1963, \mn@doi [\nat] {10.1038/199682a0}, \href
  {http://adsabs.harvard.edu/abs/1963Natur.199..682S} {199, 682}

\bibitem[\protect\citeauthoryear{{Sokolovsky}, {Kovalev}, {Pushkarev}, {Mimica}
   \& {Perucho}}{{Sokolovsky} et~al.}{2011}]{Sokolovsky2011}
{Sokolovsky} K.~V.,  {Kovalev} Y.~Y.,  {Pushkarev} A.~B.,  {Mimica} P.,
  {Perucho} M.,  2011, \mn@doi [\aap] {10.1051/0004-6361/201015772}, \href
  {http://adsabs.harvard.edu/abs/2011A%26A...535A..24S} {535, A24}

\bibitem[\protect\citeauthoryear{{Spoelstra}, {Patnaik}  \&
  {Gopal-Krishna}}{{Spoelstra} et~al.}{1985}]{Spoelstra1985}
{Spoelstra} T.~A.~T.,  {Patnaik} A.~R.,   {Gopal-Krishna} 1985, \aap, \href
  {http://adsabs.harvard.edu/abs/1985A%26A...152...38S} {152, 38}

\bibitem[\protect\citeauthoryear{{Stanghellini}, {Dallacasa}, {O'Dea}, {Baum},
  {Fanti}  \& {Fanti}}{{Stanghellini} et~al.}{2001}]{Stanghellini2001}
{Stanghellini} C.,  {Dallacasa} D.,  {O'Dea} C.~P.,  {Baum} S.~A.,  {Fanti} R.,
    {Fanti} C.,  2001, \mn@doi [\aap] {10.1051/0004-6361:20011101}, \href
  {http://adsabs.harvard.edu/abs/2001A%26A...377..377S} {377, 377}

\bibitem[\protect\citeauthoryear{{Stawarz}, {Ostorero}, {Begelman}, {Moderski},
  {Kataoka}  \& {Wagner}}{{Stawarz} et~al.}{2008}]{Stawarz2008}
{Stawarz} {\L}.,  {Ostorero} L.,  {Begelman} M.~C.,  {Moderski} R.,  {Kataoka}
  J.,   {Wagner} S.,  2008, \mn@doi [\apj] {10.1086/587781}, \href
  {http://adsabs.harvard.edu/abs/2008ApJ...680..911S} {680, 911}

\bibitem[\protect\citeauthoryear{{Swarup}, {Ananthakrishnan}, {Kapahi}, {Rao},
  {Subrahmanya}  \& {Kulkarni}}{{Swarup} et~al.}{1991}]{Swarup1991}
{Swarup} G.,  {Ananthakrishnan} S.,  {Kapahi} V.~K.,  {Rao} A.~P.,
  {Subrahmanya} C.~R.,   {Kulkarni} V.~K.,  1991, Current Science, Vol.~60,
  NO.2/JAN25, P.~95, 1991, \href
  {http://adsabs.harvard.edu/abs/1991CuSc...60...95S} {60, 95}

\bibitem[\protect\citeauthoryear{{Taylor} et~al.,}{{Taylor}
  et~al.}{2005}]{Taylor2005}
{Taylor} G.~B.,  et~al., 2005, \mn@doi [\apjs] {10.1086/430255}, \href
  {http://adsabs.harvard.edu/abs/2005ApJS..159...27T} {159, 27}

\bibitem[\protect\citeauthoryear{{Tingay} \& {Murphy}}{{Tingay} \&
  {Murphy}}{2001}]{Tingay2001}
{Tingay} S.~J.,  {Murphy} D.~W.,  2001, \mn@doi [\apj] {10.1086/318247}, \href
  {http://adsabs.harvard.edu/abs/2001ApJ...546..210T} {546, 210}

\bibitem[\protect\citeauthoryear{{Tingay} \& {de Kool}}{{Tingay} \& {de
  Kool}}{2003}]{Tingay2003}
{Tingay} S.~J.,  {de Kool} M.,  2003, \mn@doi [\aj] {10.1086/376600}, \href
  {http://adsabs.harvard.edu/abs/2003AJ....126..723T} {126, 723}

\bibitem[\protect\citeauthoryear{{Vermeulen}, {Ros}, {Kellermann}, {Cohen},
  {Zensus}  \& {van Langevelde}}{{Vermeulen} et~al.}{2003}]{Vermeulen2003}
{Vermeulen} R.~C.,  {Ros} E.,  {Kellermann} K.~I.,  {Cohen} M.~H.,  {Zensus}
  J.~A.,   {van Langevelde} H.~J.,  2003, \mn@doi [\aap]
  {10.1051/0004-6361:20021752}, \href
  {http://adsabs.harvard.edu/abs/2003A%26A...401..113V} {401, 113}

\bibitem[\protect\citeauthoryear{{V{\'e}ron-Cetty} \&
  {V{\'e}ron}}{{V{\'e}ron-Cetty} \& {V{\'e}ron}}{2010}]{Veron2010}
{V{\'e}ron-Cetty} M.-P.,  {V{\'e}ron} P.,  2010, \mn@doi [\aap]
  {10.1051/0004-6361/201014188}, \href
  {http://adsabs.harvard.edu/abs/2010A%26A...518A..10V} {518, A10}

\bibitem[\protect\citeauthoryear{{Walker}, {Romney}  \& {Benson}}{{Walker}
  et~al.}{1994}]{Walker1994}
{Walker} R.~C.,  {Romney} J.~D.,   {Benson} J.~M.,  1994, in {Zensus} J.~A.,
  {Kellermann} K.~I.,  eds, Compact Extragalactic Radio Sources. p.~121

\bibitem[\protect\citeauthoryear{{Wayth} et~al.,}{{Wayth}
  et~al.}{2015}]{Wayth2015}
{Wayth} R.~B.,  et~al., 2015, \mn@doi [\pasa] {10.1017/pasa.2015.26}, \href
  {http://adsabs.harvard.edu/abs/2015PASA...32...25W} {32, e025}

\bibitem[\protect\citeauthoryear{{Wilkinson}, {Polatidis}, {Readhead}, {Xu}  \&
  {Pearson}}{{Wilkinson} et~al.}{1994}]{Wilkinson1994}
{Wilkinson} P.~N.,  {Polatidis} A.~G.,  {Readhead} A.~C.~S.,  {Xu} W.,
  {Pearson} T.~J.,  1994, \mn@doi [\apjl] {10.1086/187518}, \href
  {http://adsabs.harvard.edu/abs/1994ApJ...432L..87W} {432, L87}

\bibitem[\protect\citeauthoryear{{Xu}, {Readhead}, {Pearson}, {Polatidis}  \&
  {Wilkinson}}{{Xu} et~al.}{1995}]{Xu1995}
{Xu} W.,  {Readhead} A.~C.~S.,  {Pearson} T.~J.,  {Polatidis} A.~G.,
  {Wilkinson} P.~N.,  1995, \mn@doi [\apjs] {10.1086/192189}, \href
  {http://adsabs.harvard.edu/abs/1995ApJS...99..297X} {99, 297}

\bibitem[\protect\citeauthoryear{{Zovaro}, {Sharp}, {Nesvadba}, {Bicknell},
  {Mukherjee}, {Wagner}, {Groves}  \& {Krishna}}{{Zovaro}
  et~al.}{2019}]{Zovaro2019}
{Zovaro} H. R.~M.,  {Sharp} R.,  {Nesvadba} N. P.~H.,  {Bicknell} G.~V.,
  {Mukherjee} D.,  {Wagner} A. e.~Y.,  {Groves} B.,   {Krishna} S.,  2019,
  \mn@doi [\mnras] {10.1093/mnras/stz233}, \href
  {https://ui.adsabs.harvard.edu/abs/2019MNRAS.484.3393Z} {484, 3393}

\bibitem[\protect\citeauthoryear{{de Bruyn} et~al.,}{{de Bruyn}
  et~al.}{2000}]{deBruyn2000}
{de Bruyn} G.,  et~al., 2000, VizieR Online Data Catalog, \href
  {http://adsabs.harvard.edu/abs/2000yCat.8062....0D} {8062}

\bibitem[\protect\citeauthoryear{{de Kool} \& {Begelman}}{{de Kool} \&
  {Begelman}}{1989}]{deKool1989}
{de Kool} M.,  {Begelman} M.~C.,  1989, \mn@doi [\nat] {10.1038/338484a0},
  \href {http://adsabs.harvard.edu/abs/1989Natur.338..484D} {338, 484}

\bibitem[\protect\citeauthoryear{{de Vries}, {Barthel}  \& {O'Dea}}{{de Vries}
  et~al.}{1997}]{deVries1997}
{de Vries} W.~H.,  {Barthel} P.~D.,   {O'Dea} C.~P.,  1997, \aap, \href
  {http://adsabs.harvard.edu/abs/1997A%26A...321..105D} {321, 105}

\makeatother
\end{thebibliography}
\clearpage

\bsp	
\label{lastpage}
\end{document}


\begin{strip}
\thispagestyle{empty}
\begin{center}	
\vspace*{-3.0cm} 
{\textbf{ONLINE MATERIAL}}\\
{GMRT observations of a first sample of \lq Extremely Inverted Spectrum Extragalactic Radio Sources (EISERS)\rq~candidates in the Northern sky}\\
\vspace*{0.6cm} \small{{Mukul Mhaskey$^{1,2}$, Gopal-Krishna$^{3}$, Surajit Paul$^{1}$, Pratik Dabhade$^{4,5}$}\\
 {Sameer Salunkhe$^{1}$, Shubham Bhagat$^{1}$ and Abhijit Bendre$^{4}$}}\\
 
\vspace*{0.3cm} \it \tiny{$^{1}$Department of Physics, Savitribai Phule Pune Unversity, Ganeshkhind, Pune 411007, India\\
 $^{2}$Th{\"u}ringer Landessternwarte, Sternwarte 5, D-07778 Tautenburg, Germany \\
 $^{3}$Aryabhatta Research Institute of Observational Sciences (ARIES), Manora Peak, Nainital $-$ 263129, India\\
 $^{4}$Inter University Centre for Astronomy and Astrophysics (IUCAA), Pune 411007, India\\
 $^{5}$Leiden Observatory, Leiden University, Niels Bohrweg 2, 2333 CA, Leiden, Netherlands}\\
\end{center}
\end{strip}


\begin{figure}
\includegraphics[width=3.0in]{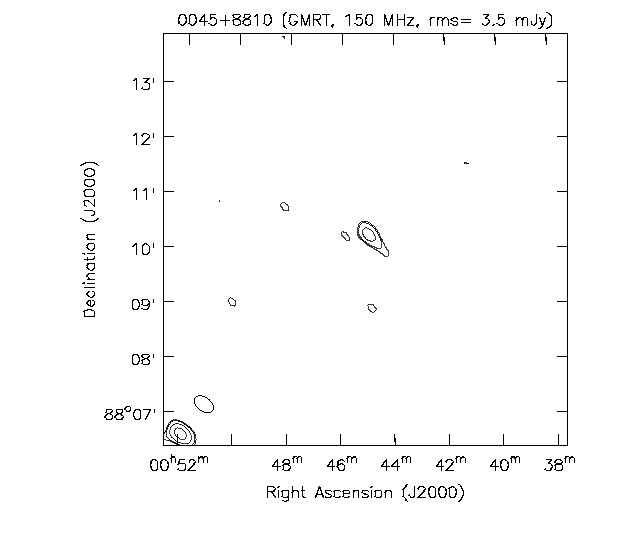} 
\includegraphics[width=3.0in]{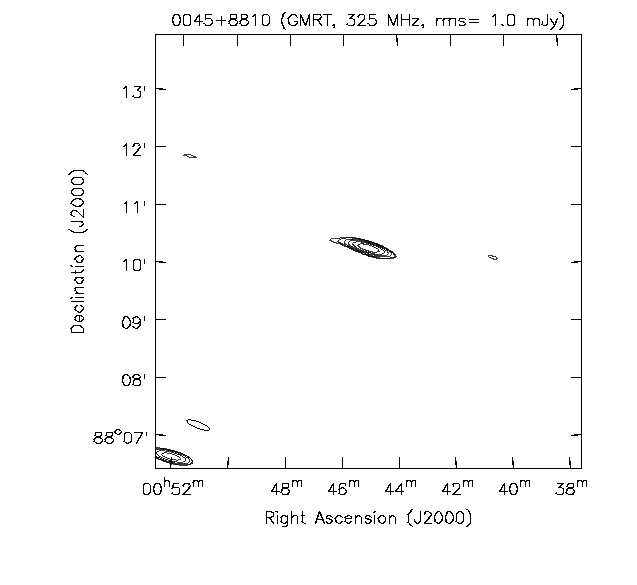} 
\caption{\scriptsize{GMRT contour maps of J0045$+$8810 at 150 MHz and 325 MHz, respectively. The contour levels are 3,4,8,16,32,64 \& 128 with the unit contour level at 150 MHz at 3.5 mJy and 1.0 mJy at 325 MHz. The beam size is $24~\times~15^{"}$ (PA= $53^{\circ}$) and $25~\times~06{"}$ (PA= $69^{\circ}$) at 150 MHz and 325 MHz respectively. The target source lies at the centre of the map.}}   
\end{figure}

\begin{figure}
\includegraphics[width=3.0in]{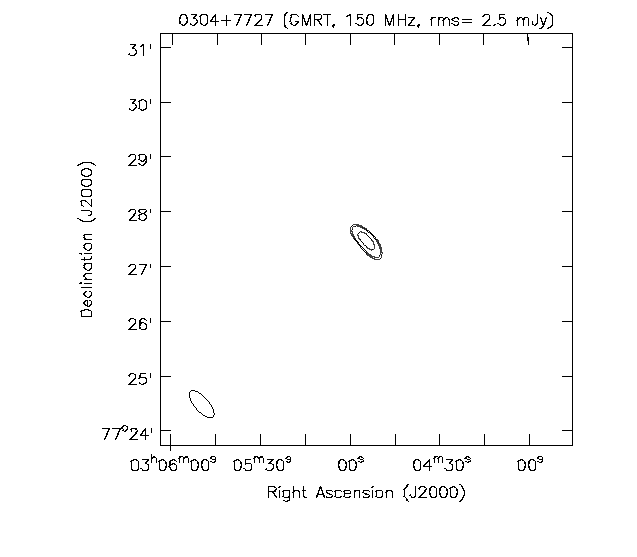} 
\includegraphics[width=3.0in]{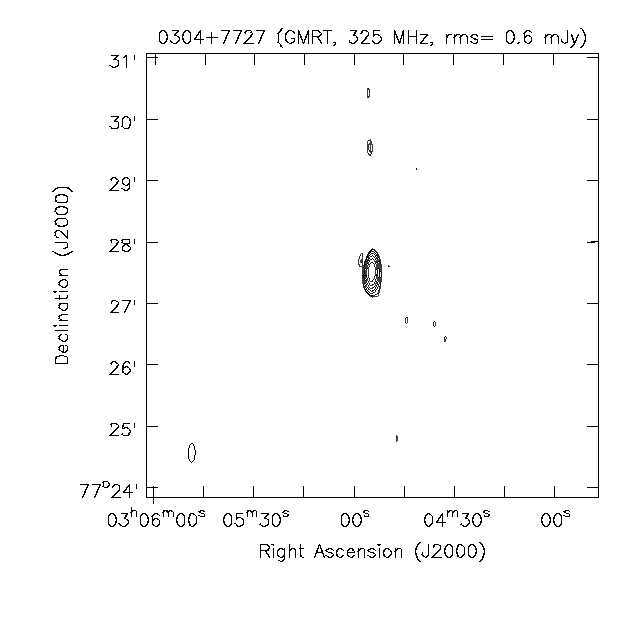} 
\caption{\scriptsize{GMRT contour maps of J0304$+$7727 at 150 MHz and 325 MHz, respectively. The contour levels are 3,4,8,16,32,64 \& 128 with the unit contour level at 150 MHz at 2.5 mJy and 0.6 mJy at 325 MHz. The beam size is $38~\times~15^{"}$ (PA= $41^{\circ}$) and $18~\times~07{"}$ (PA= $-01^{\circ}$) at 150 MHz and 325 MHz respectively. The target source lies at the centre of the map.}}  
\end{figure}

\begin{figure}
\includegraphics[width=3.0in]{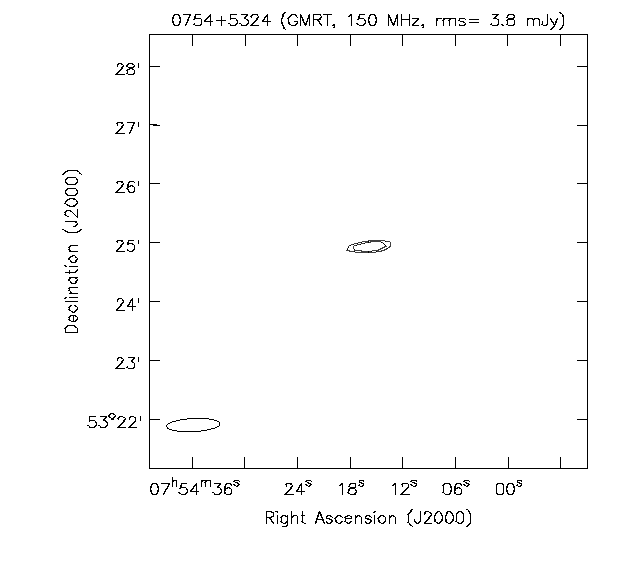} 
\includegraphics[width=3.0in]{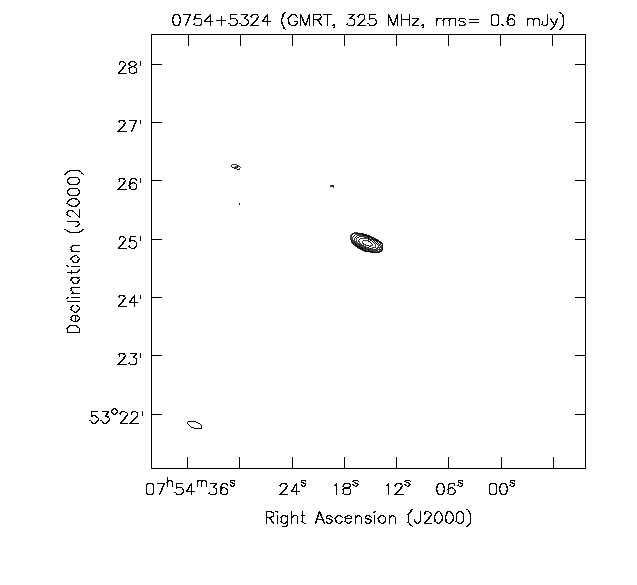} 
\caption{\scriptsize{GMRT contour maps of J0754$+$5324 at 150 MHz and 325 MHz, respectively. The contour levels are 3,4,8,16,32,64 \& 128 with the unit contour level at 150 MHz at 3.8 mJy and 0.6 mJy at 325 MHz. The beam size is $55~\times~13^{"}$ (PA= $-87{\circ}$) and $15~\times~07{"}$ (PA= $70^{\circ}$) at 150 MHz and 325 MHz respectively. The target source lies at the centre of the map.}}   
\end{figure}

\begin{figure}
\includegraphics[width=3.0in]{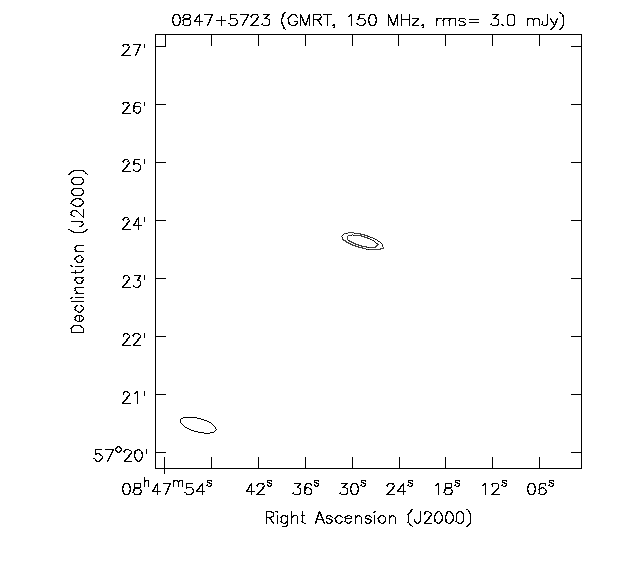} 
\includegraphics[width=3.0in]{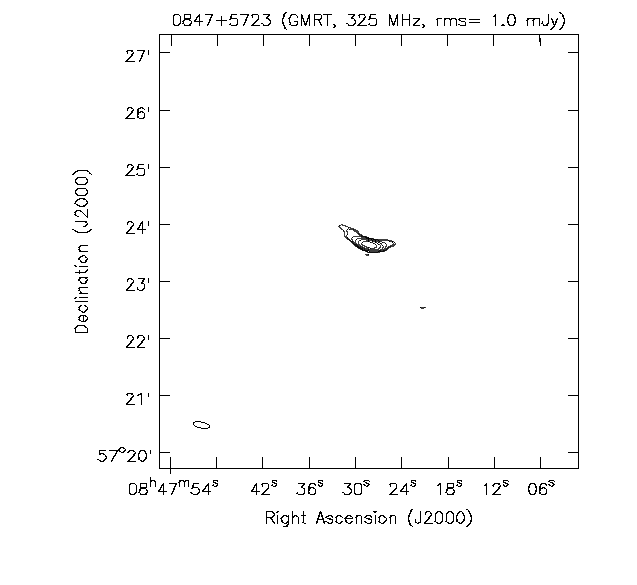} 
\caption{\scriptsize{GMRT contour maps of J0847$+$5723 at 150 MHz and 325 MHz, respectively. The contour levels are 3,4,8,16,32,64 \& 128 with the unit contour level at 150 MHz at 3.0 mJy and 1.0 mJy at 325 MHz. The beam size is $38~\times~14^{"}$ (PA= $75^{\circ}$) and $18~\times~06{"}$ (PA= $78^{\circ}$) at 150 MHz and 325 MHz respectively. The target source lies at the centre of the map}}  
\end{figure}

\begin{figure}
\includegraphics[width=3.0in]{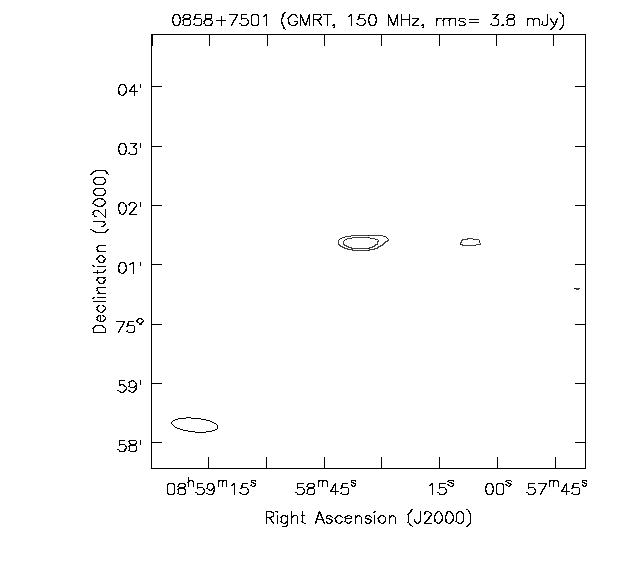} 
\includegraphics[width=3.0in]{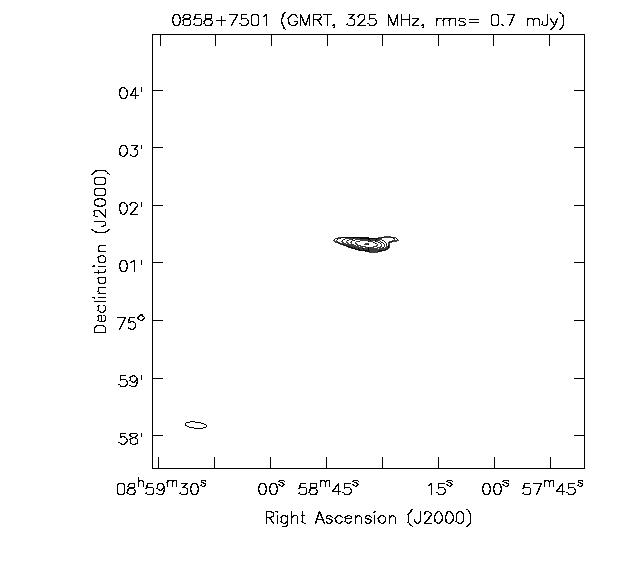} 
\caption{\scriptsize{GMRT contour maps of J0858$+$7501 at 150 MHz and 325 MHz, respectively. The contour levels are 3,4,8,16,32,64 \& 128 with the unit contour level at 150 MHz at 3.8 mJy and 0.7 mJy at 325 MHz. The beam size is $47~\times~14^{"}$ (PA= $85^{\circ}$) and $22~\times~06{"}$ (PA= $84^{\circ}$) at 150 MHz and 325 MHz respectively. The target source lies at the centre of the map.}}  
\end{figure}

\begin{figure}
\includegraphics[width=3.0in]{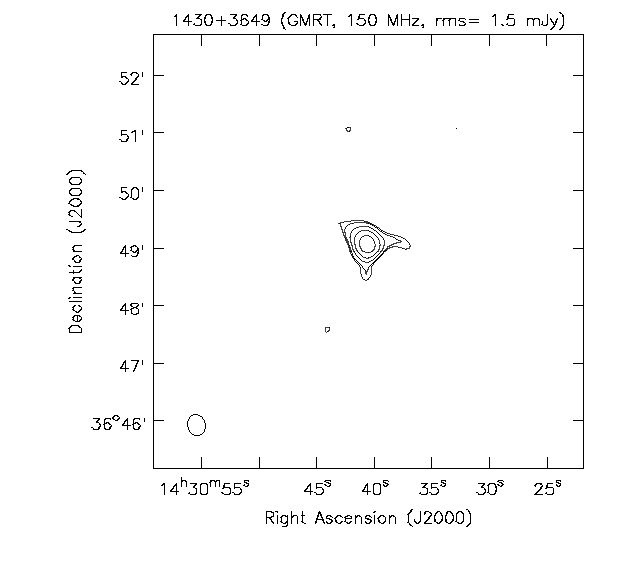} 
\includegraphics[width=3.0in]{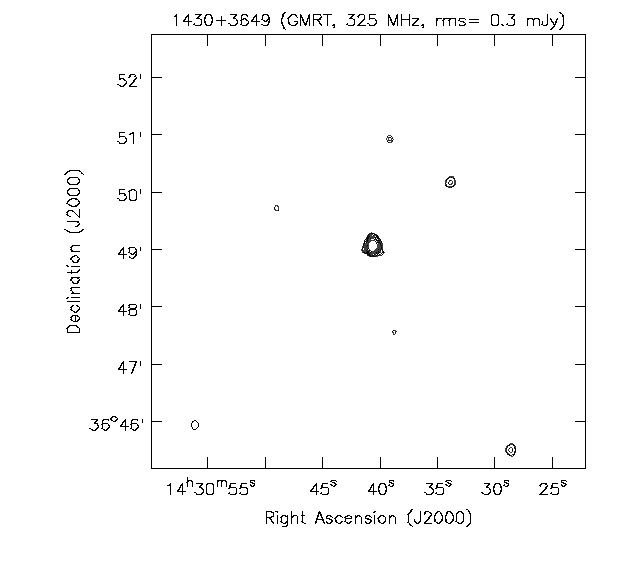} 
\caption{\scriptsize{GMRT contour maps of J1430$+$3649 at 150 MHz and 325 MHz, respectively. The contour levels are 3,4,8,16,32,64 \& 128 with the unit contour level at 150 MHz at 1.5 mJy and 0.3 mJy at 325 MHz. The beam size is $22~\times~18^{"}$ (PA= $14^{\circ}$) and $09~\times~07{"}$ (PA= $02^{\circ}$) at 150 MHz and 325 MHz respectively. The target source lies at the centre of the map.}}  
\end{figure}

\begin{figure}
\includegraphics[width=3.0in]{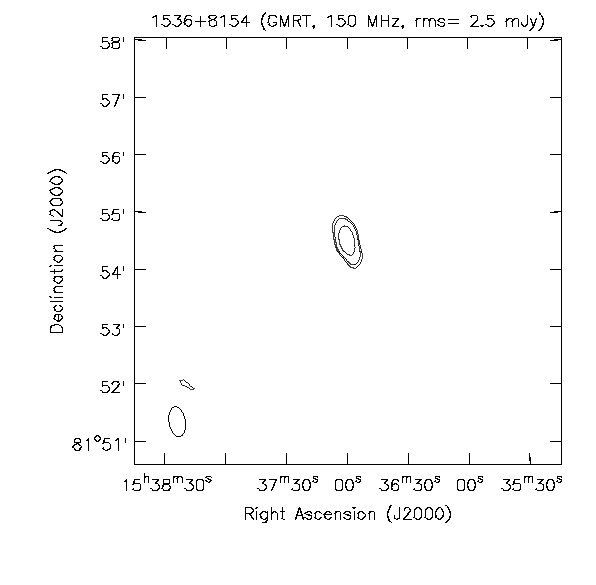} 
\includegraphics[width=3.0in]{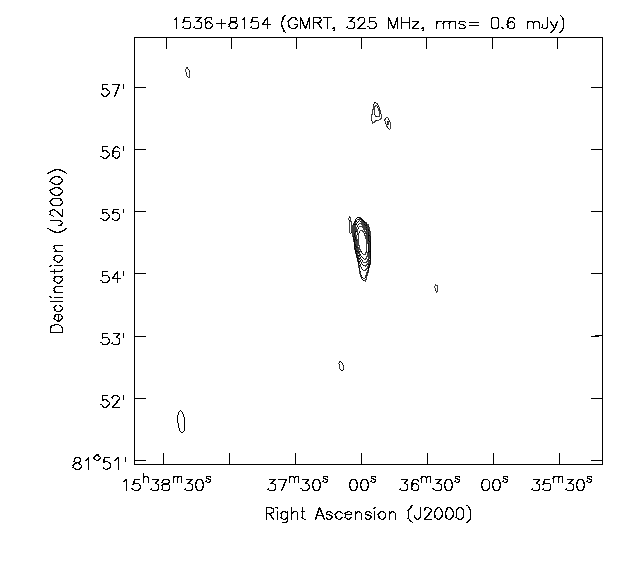} 
\caption{\scriptsize{GMRT contour maps of J1536$+$8154 at 150 MHz and 325 MHz, respectively. The contour levels are 3,4,8,16,32,64 \& 128 with the unit contour level at 150 MHz at 2.5 mJy and 0.6 mJy at 325 MHz. The beam size is $32~\times~17^{"}$ (PA= $10^{\circ}$) and $21~\times~06{"}$ (PA= $05^{\circ}$) at 150 MHz and 325 MHz respectively. The target source lies at the centre of the map.}}  
\end{figure}

\begin{figure}
\includegraphics[width=3.0in]{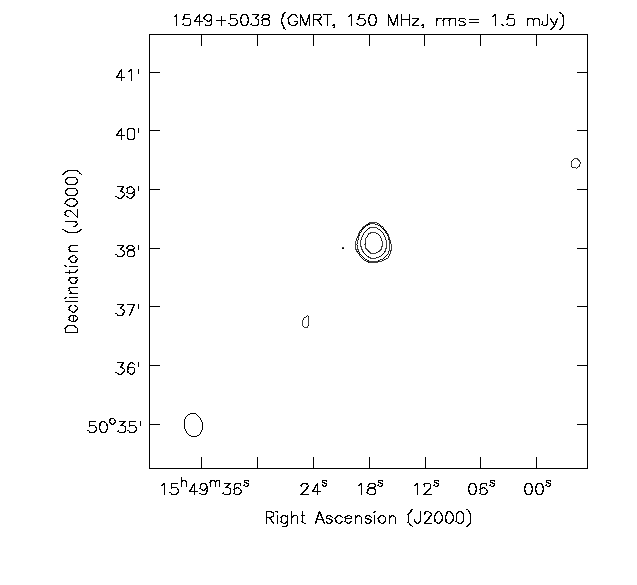} 
\includegraphics[width=3.0in]{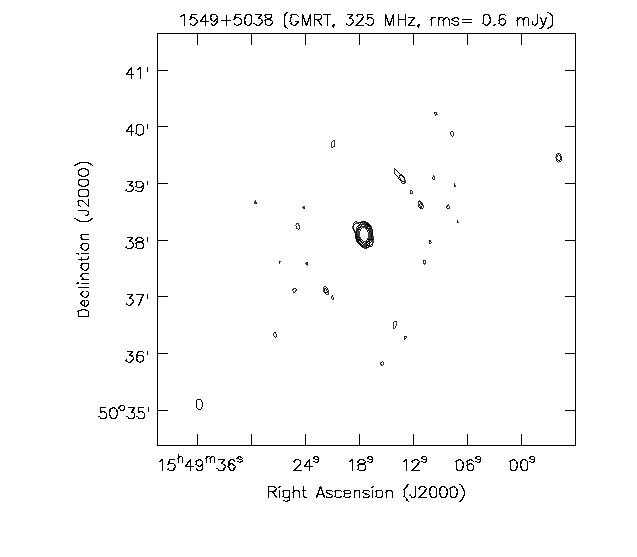} 
\caption{\scriptsize{GMRT contour maps of J1549$+$5038 at 150 MHz and 325 MHz, respectively. The contour levels are 3,4,8,16,32,64 \& 128 with the unit contour level at 150 MHz at 1.5 mJy and 0.6 mJy at 325 MHz. The beam size is $24~\times~19^{"}$ (PA= $09^{\circ}$) and $11~\times~07{"}$ (PA= $04^{\circ}$) at 150 MHz and 325 MHz respectively. The target source lies at the centre of the map.}} 
\end{figure}

\begin{figure}
\includegraphics[width=3.0in]{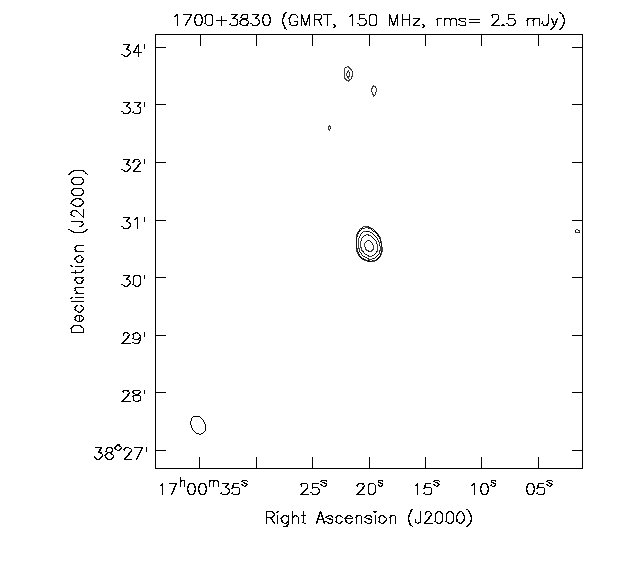} 
\includegraphics[width=3.0in]{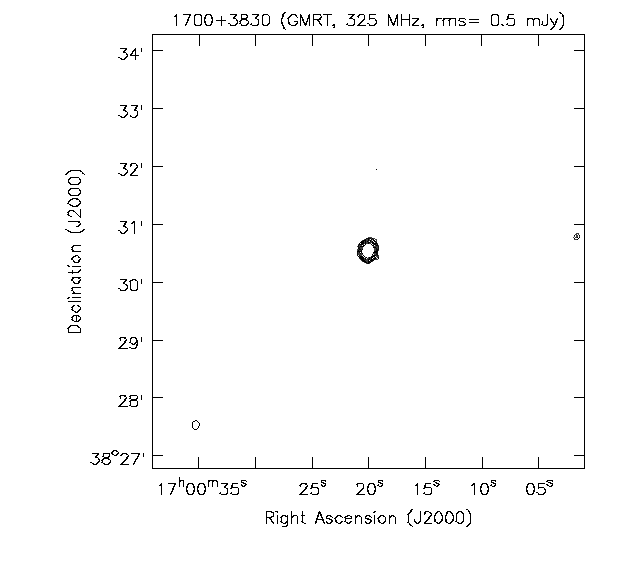} 
\caption{\scriptsize{GMRT contour maps of J1700$+$3830 at 150 MHz and 325 MHz, respectively. The contour levels are 3,4,8,16,32,64 \& 128 with the unit contour level at 150 MHz at 2.5 mJy and 0.5 mJy at 325 MHz. The beam size is $20~\times~14^{"}$ (PA= $26^{\circ}$) and $09~\times~07{"}$ (PA= $-12^{\circ}$) at 150 MHz and 325 MHz respectively. The target source lies at the centre of the map.}}  
\end{figure}

\begin{figure}
\includegraphics[width=3.0in]{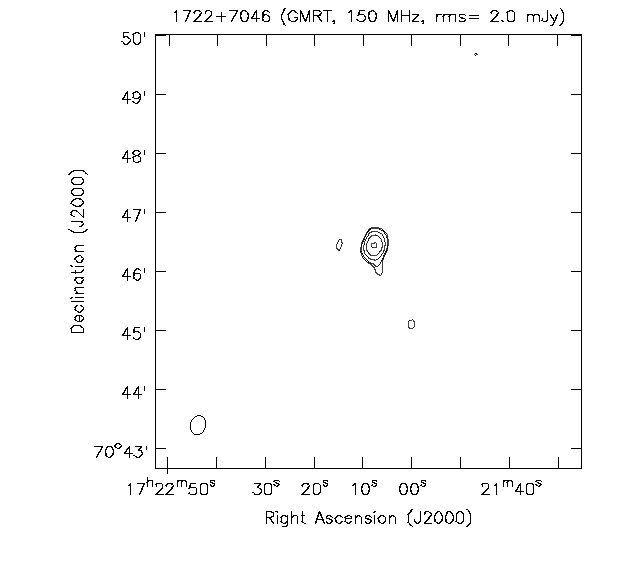} 
\includegraphics[width=3.0in]{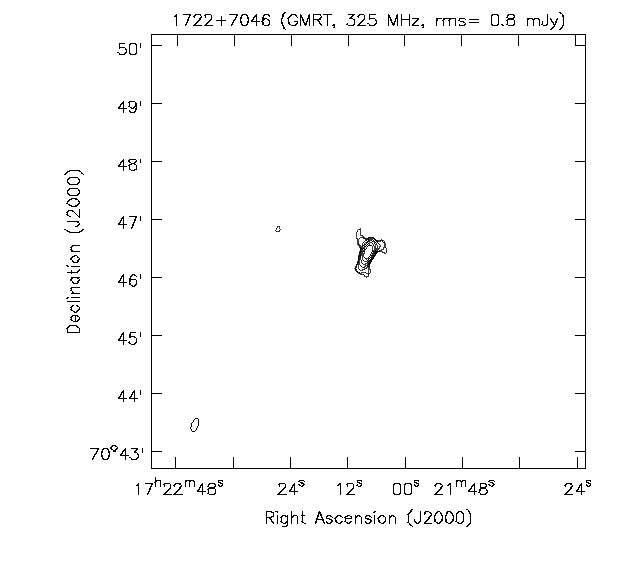} 
\caption{\scriptsize{GMRT contour maps of J1722$+$7046 at 150 MHz and 325 MHz, respectively. The contour levels are 3,4,8,16,32,64 \& 128 with the unit contour level at 150 MHz at 2.0 mJy and 0.8 mJy at 325 MHz. The beam size is $20~\times~16^{"}$ (PA= $-14^{\circ}$) and $14~\times~07{"}$ (PA= $-22^{\circ}$) at 150 MHz and 325 MHz respectively. The target source lies at the centre of the map.}}  
\end{figure}

\begin{figure}
\includegraphics[width=3.0in]{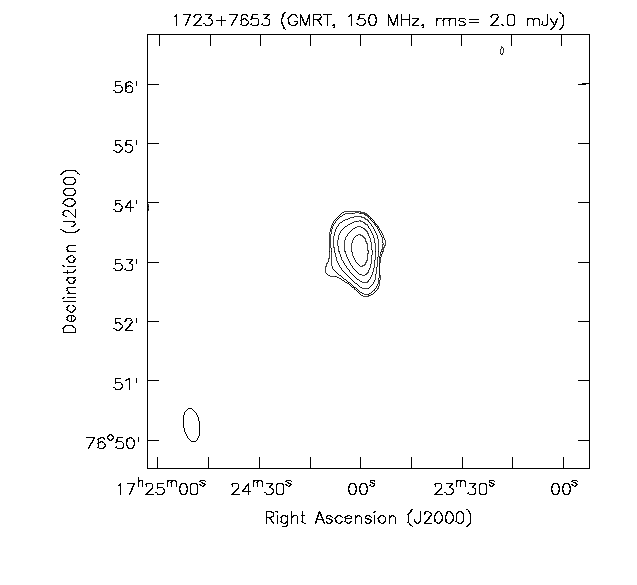} 
\includegraphics[width=3.0in]{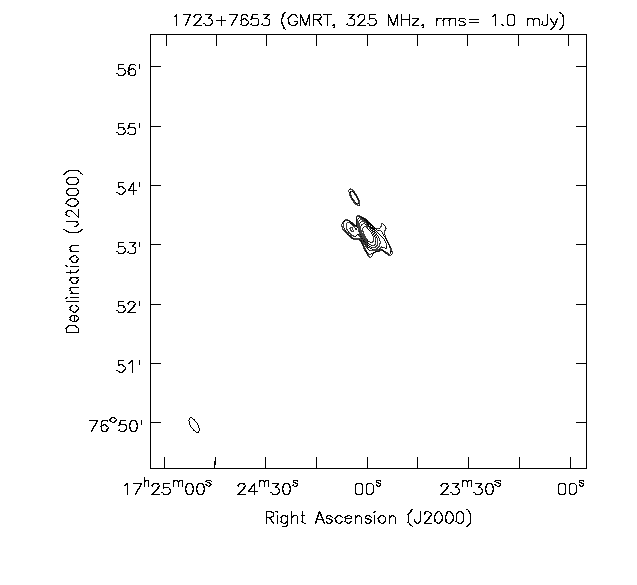} 
\caption{\scriptsize{GMRT contour maps of J1723$+$7653 at 150 MHz and 325 MHz, respectively. The contour levels are 3,4,8,16,32,64 \& 128 with the unit contour level at 150 MHz at 2.0 mJy and 1.0 mJy at 325 MHz. The beam size is $34~\times~16^{"}$ (PA= $08^{\circ}$) and $17~\times~07{"}$ (PA= $30^{\circ}$) at 150 MHz and 325 MHz respectively. The target source lies at the centre of the map.}} 
\end{figure}

\begin{figure}
\includegraphics[width=3.0in]{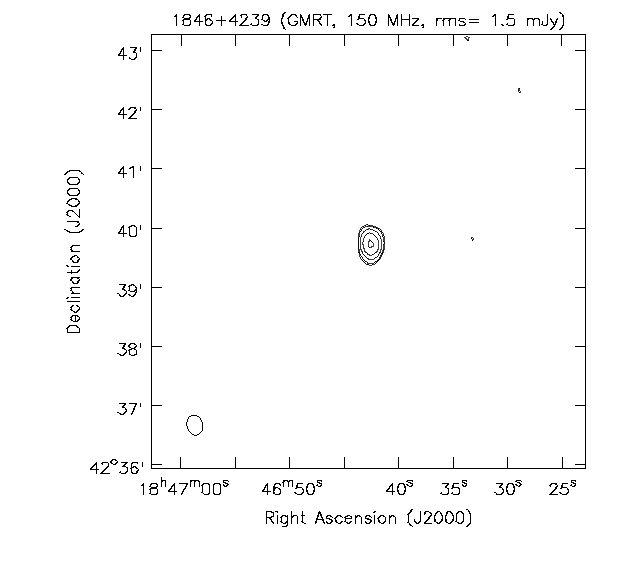} 
\includegraphics[width=3.0in]{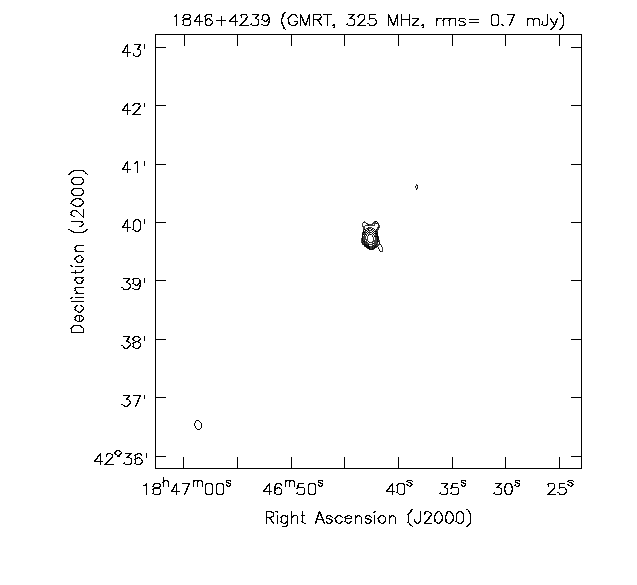} 
\caption{\scriptsize{GMRT contour maps of J1846$+$4239 at 150 MHz and 325 MHz, respectively. The contour levels are 3,4,8,16,32,64 \& 128 with the unit contour level at 150 MHz at 1.5 mJy and 0.7 mJy at 325 MHz. The beam size is $21~\times~16^{"}$ (PA= $13^{\circ}$) and $09~\times~07{"}$ (PA= $08^{\circ}$) at 150 MHz and 325 MHz respectively. The target source lies at the centre of the map.}} 
\end{figure}

\begin{figure}
\includegraphics[width=3.0in]{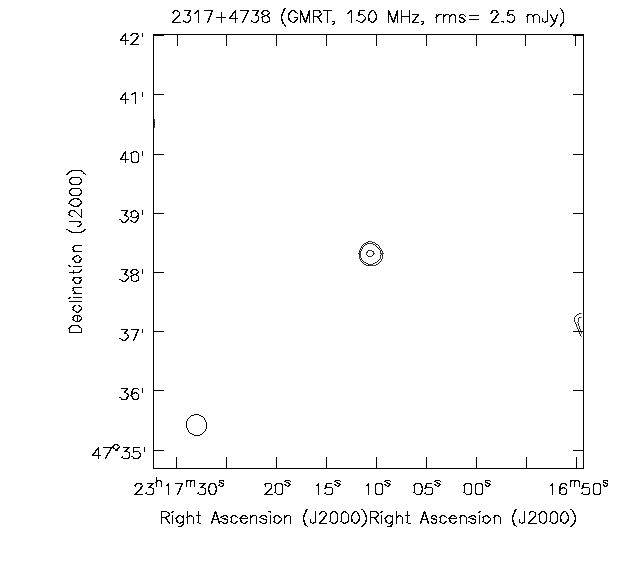} 
\includegraphics[width=3.0in]{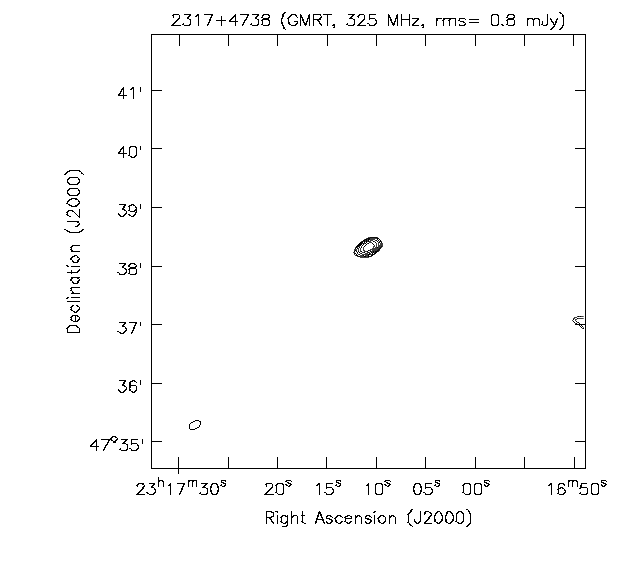} 
\caption{\scriptsize{GMRT contour maps of J2317$+$4738 at 150 MHz and 325 MHz, respectively. The contour levels are 3,4,8,16,32,64 \& 128 with the unit contour level at 150 MHz at 2.5 mJy and 0.8 mJy at 325 MHz. The beam size is $21~\times~20^{"}$ (PA= $23^{\circ}$) and $13~\times~08{"}$ (PA= $-62^{\circ}$) at 150 MHz and 325 MHz respectively. The target source lies at the centre of the map.}}  
\end{figure}

\clearpage

\begin{figure*}[h]
\centering
\includegraphics[scale=.5]{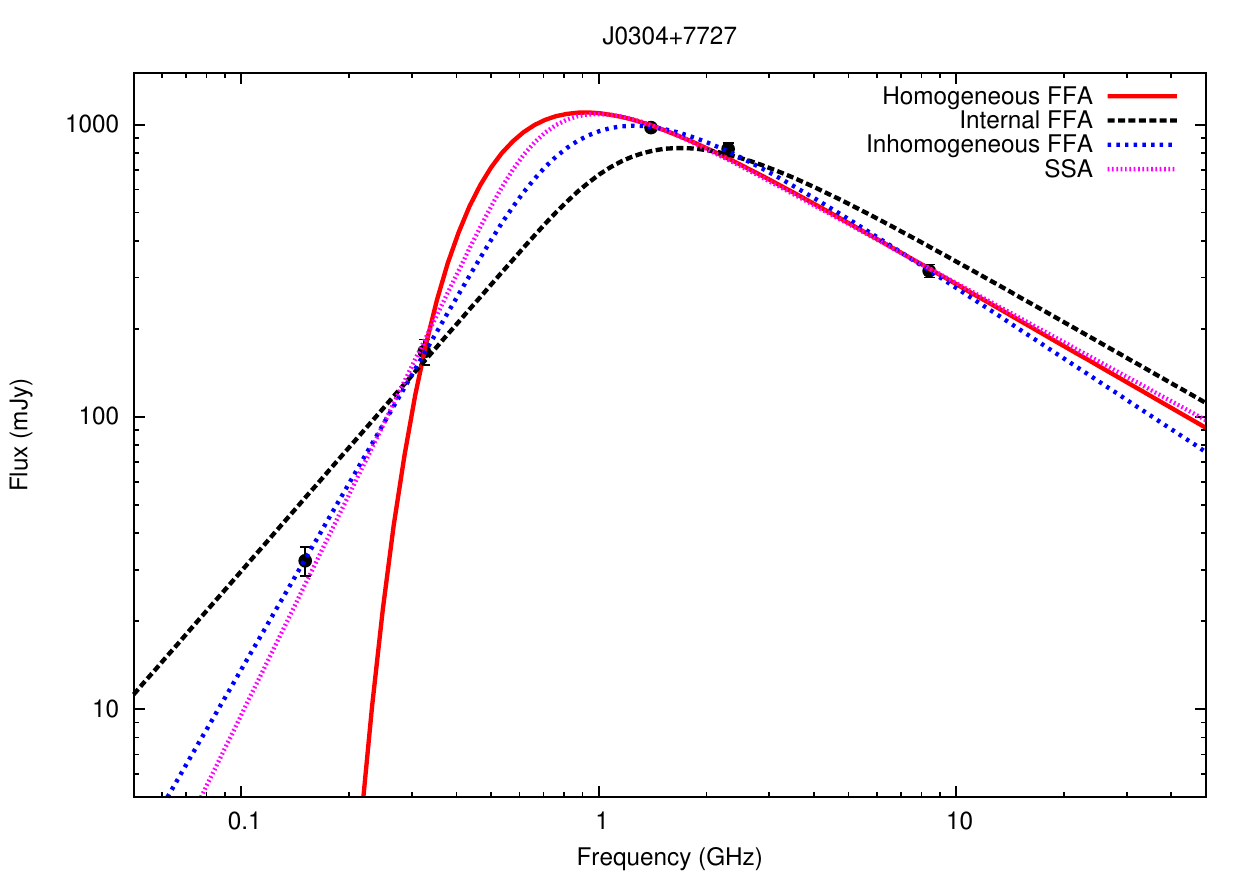} 
\includegraphics[scale=.5]{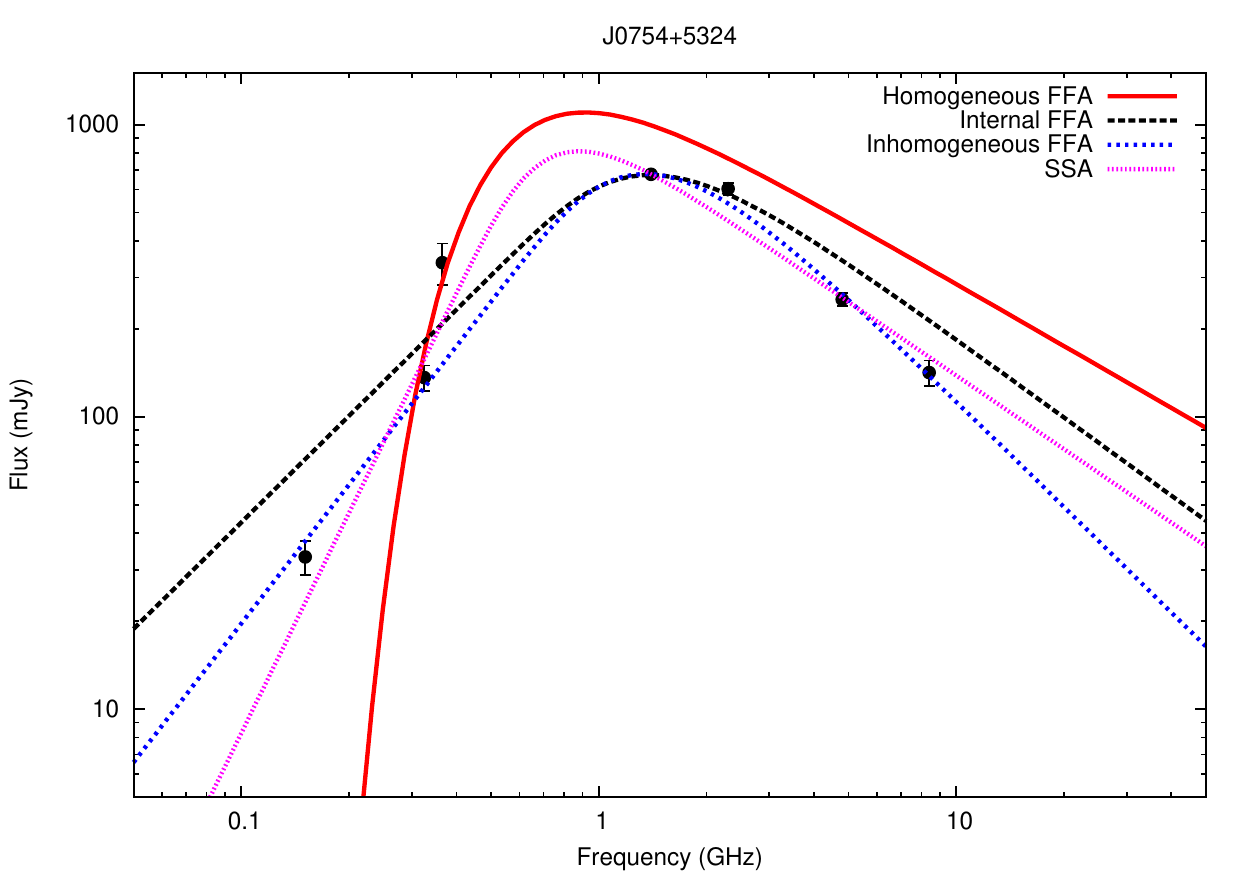} 
\includegraphics[scale=.5]{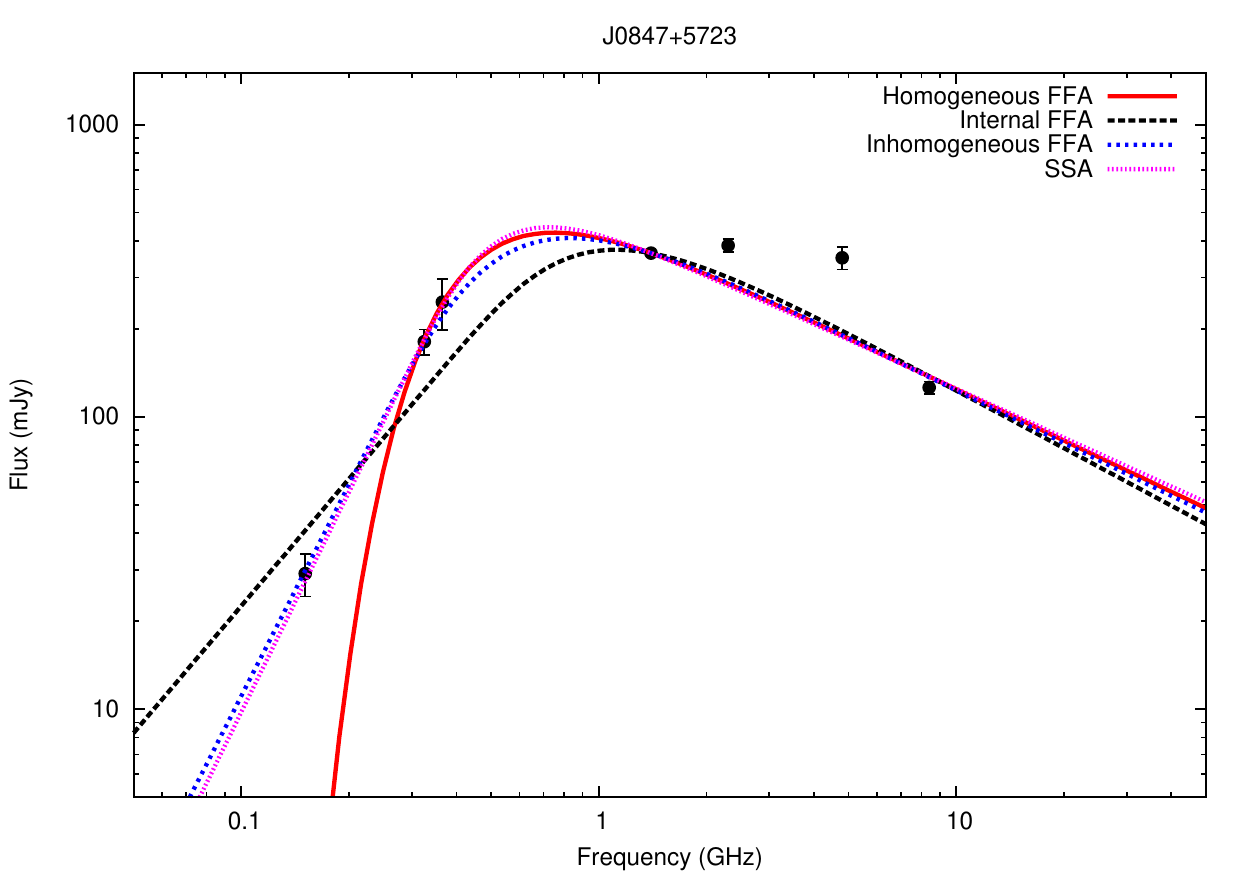} 
\includegraphics[scale=.5]{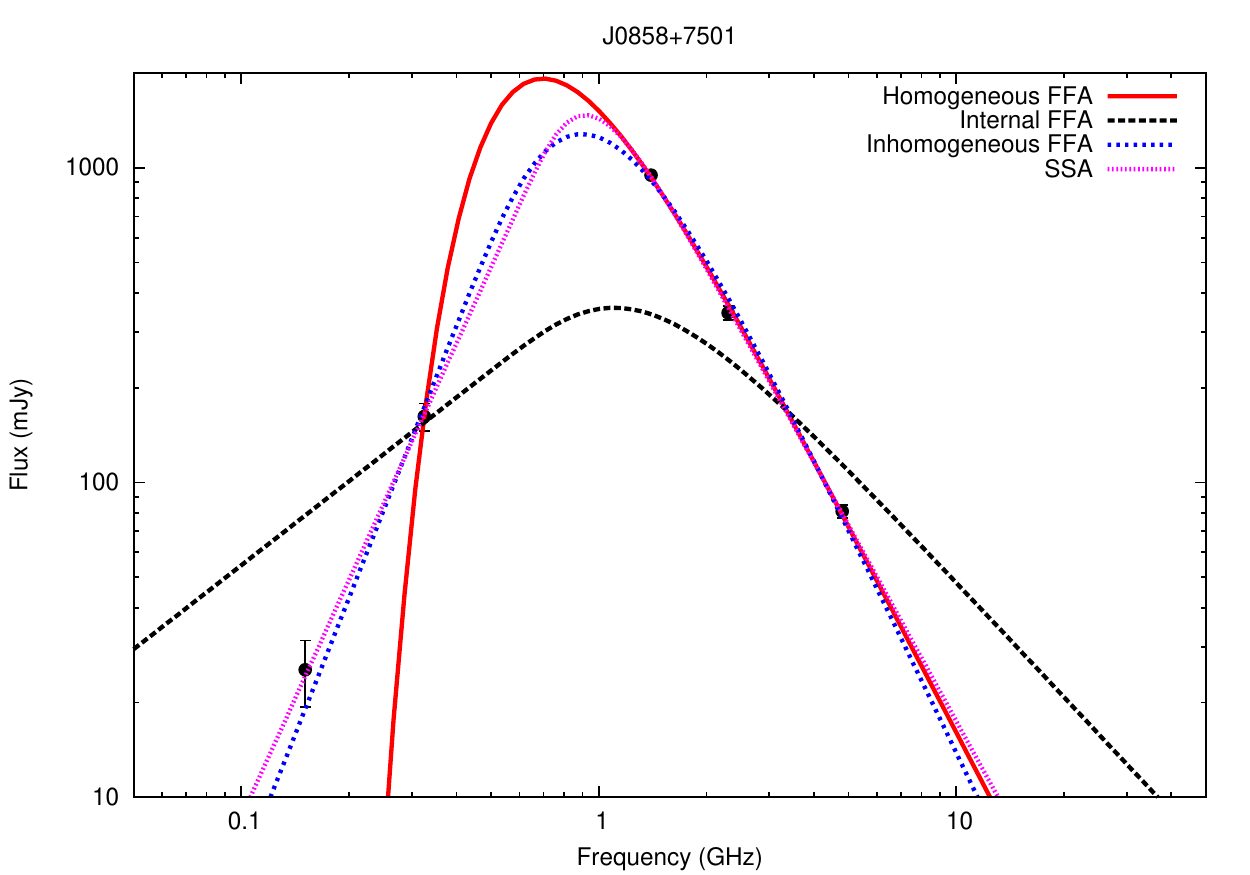} 
\includegraphics[scale=.5]{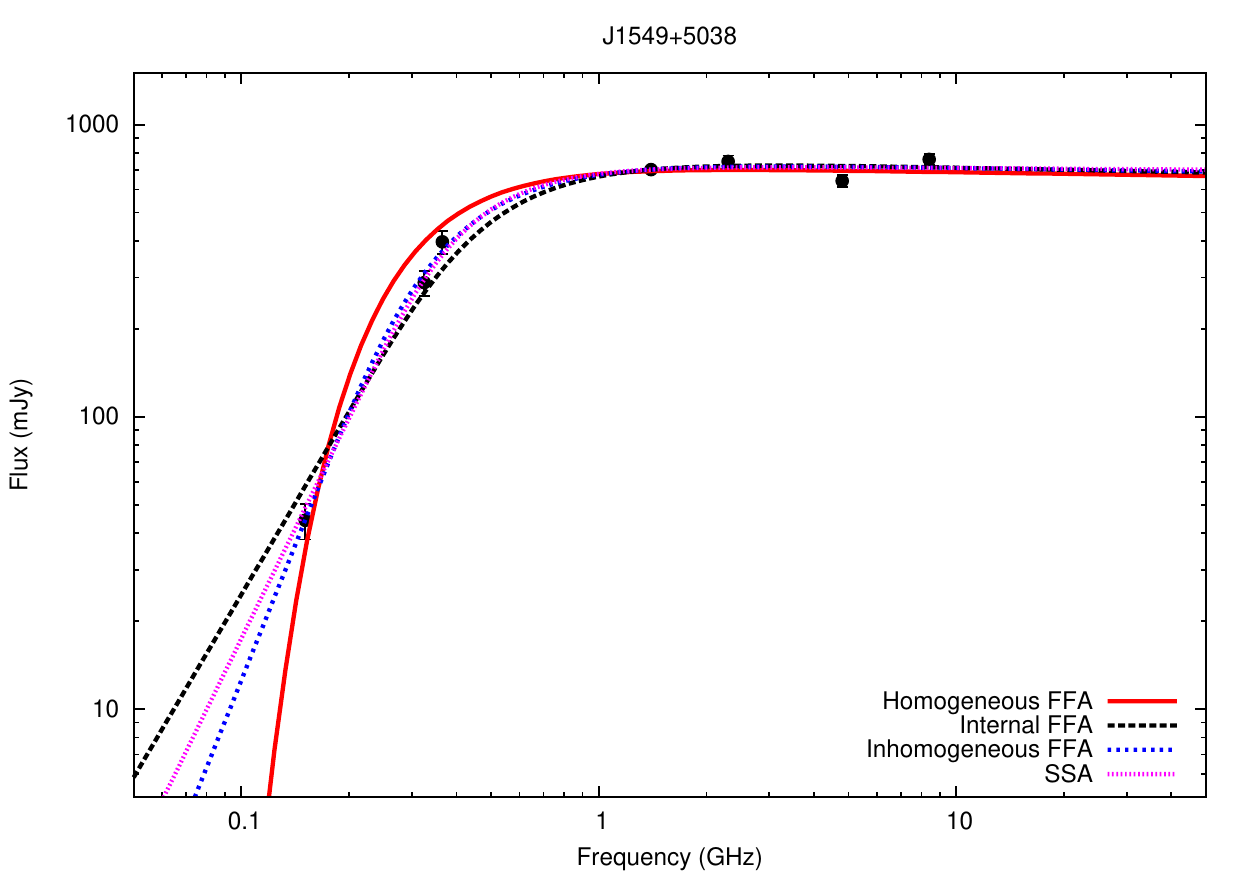} 
\includegraphics[scale=.5]{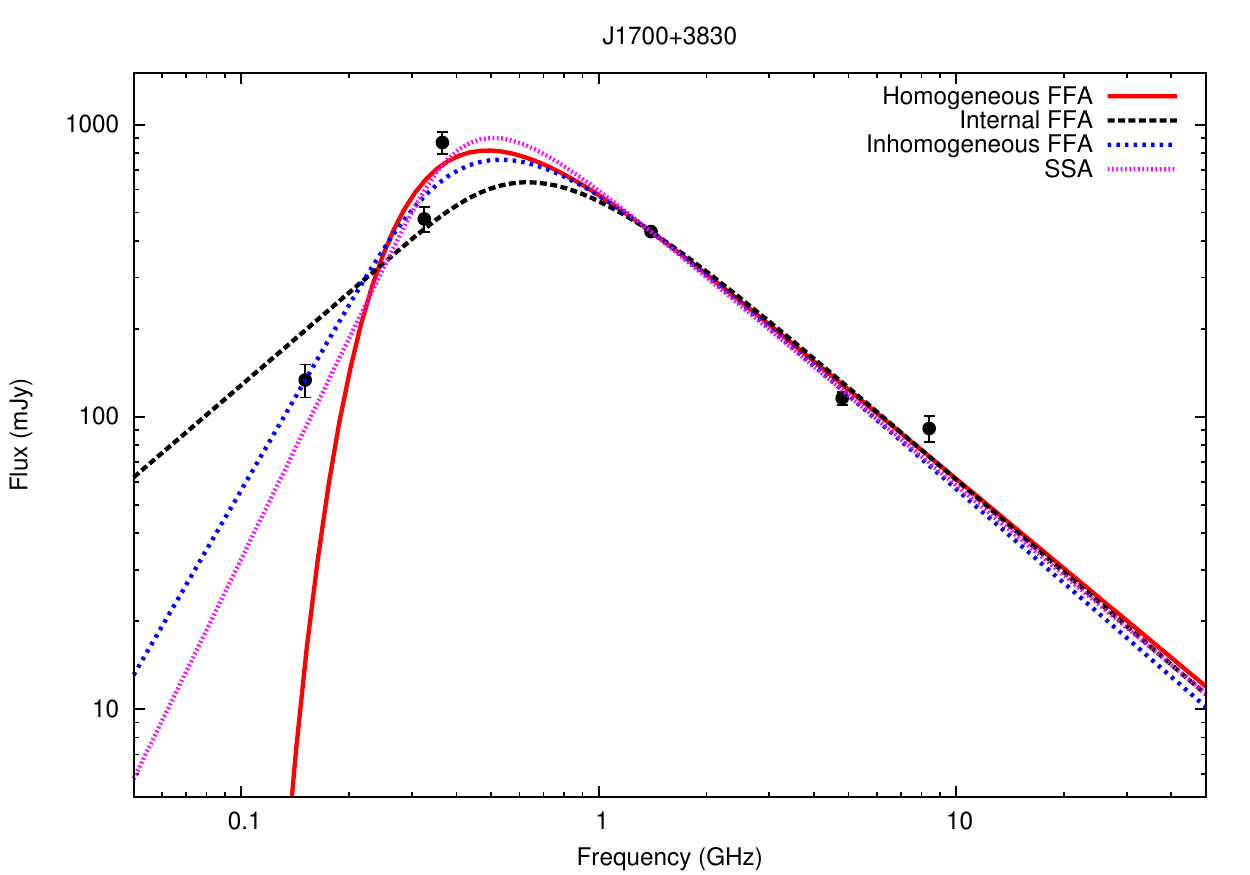} 
\includegraphics[scale=.5]{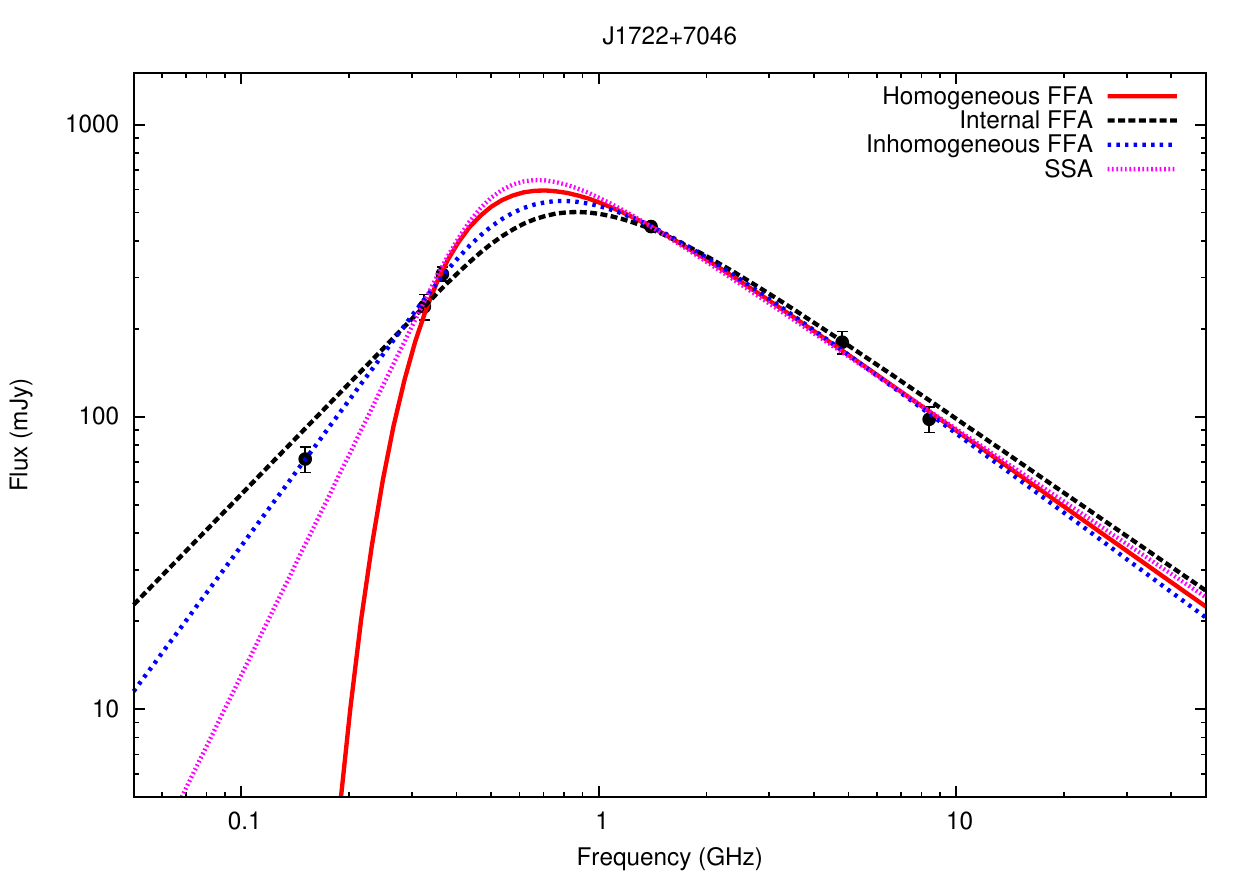}
\includegraphics[scale=.5]{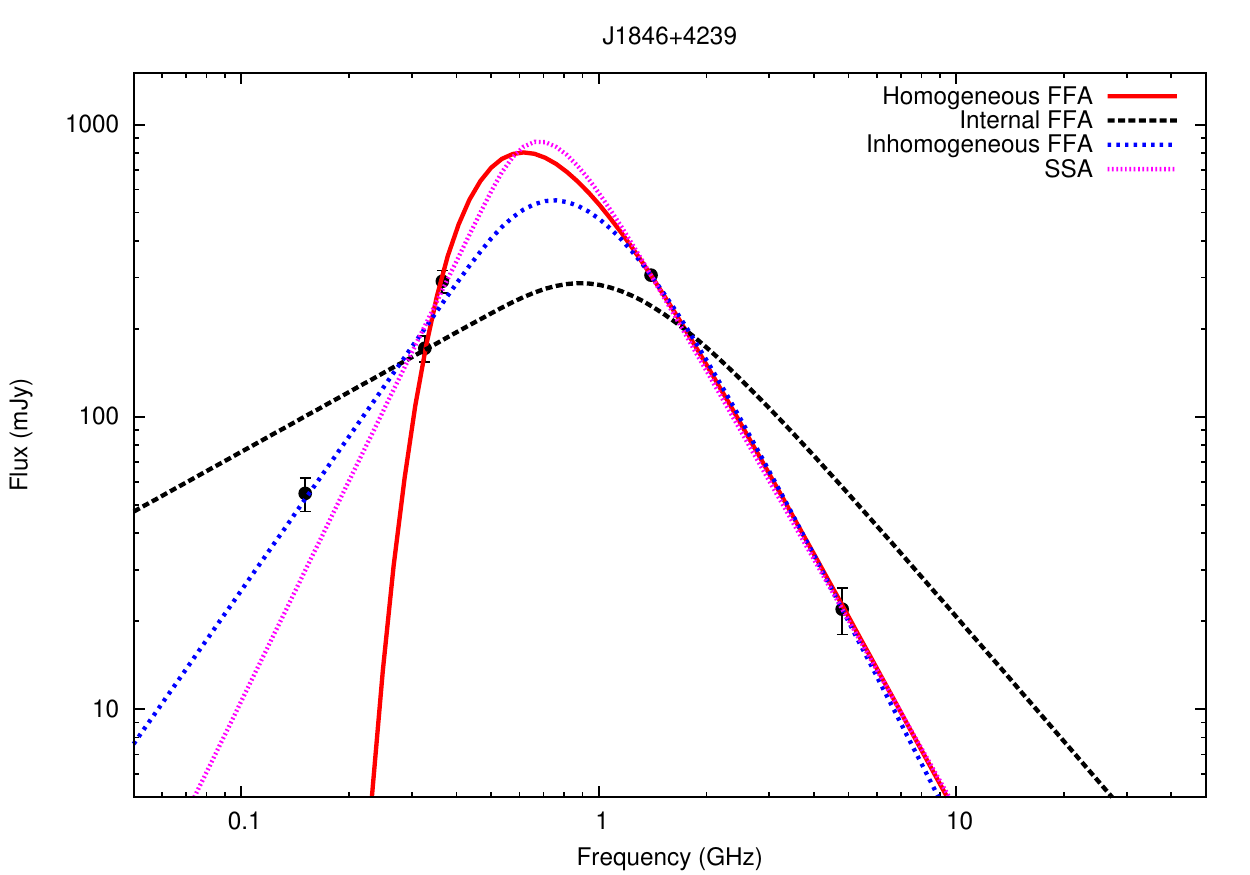}  
\caption{\scriptsize{Different absorption models fitted to the spectral energy distribution of the observed EISERS candidates.}}
\label{fig:spec_all_detected}
\end{figure*}

\clearpage